\documentclass[12pt]{aastex}

\usepackage{emulateapj5}
\usepackage{psfig}

\newcommand{\flam}{ergs~cm$^{-2}$~s$^{-1}$~\AA$^{-1}$}
\newcommand{\msun}{\mbox{$\:M_{\sun}$}}

\newcommand{\expu}[3]{\mbox{\rm $#1 \times 10^{#2} \rm\:#3$}}

\begin{document}

\submitted{To appear in the Astrophysical Journal.}

\title{Accretion and Outflow in Interacting Binary Systems:
FUSE Observations of the Novalike Cataclysmic Variable,
UX~Ursae~Majoris
}

\author{Cynthia S.\ Froning, Knox S.\ Long}
\email{froning@stsci.edu, long@stsci.edu}
\affil{Space Telescope Science Institute, \\ 3700 San Martin Drive,
Baltimore, MD 21218}
\and
\author{Christian Knigge}
\email{christian@astro.soton.ac.uk}
\affil{Department of Physics \& Astronomy,\\ University of Southampton, \\
Southampton SO17 1BJ UK}

\begin{abstract}
We present far-ultraviolet (905 -- 1182~\AA), time-series spectroscopy of
the eclipsing, novalike cataclysmic variable, UX~UMa, acquired with FUSE.
The time-averaged spectrum is complex and is dominated by overlapping
spectral features.  The most prominent features are emission lines of
\ion{C}{3}, \ion{N}{3}, \ion{N}{4}, and \ion{O}{6}.  They are broad (FWHM
$\geq$ 1800~km~s$^{-1}$) and double-peaked with a central absorption at
zero velocity.  During eclipse, the spectrum is simpler: the emission lines
remain bright, but the absorption components of the lines and the weaker
features between the emission lines disappear entirely, leaving a flat
continuum. This behavior is also evident in GHRS (1149 -- 1660~\AA) spectra
that we retrieved from the HST archive. The FUV spectra show flickering on
time scales of several minutes.  The flickering is seen primarily in the
continuum and/or the weaker lines rather than in the prominent emission
lines. The orbital light curve has a dip in the FUV flux between orbital
phases 0.45 -- 0.65, similar to a pre-eclipse dip detected in HST
observations.  The EWs of the line absorption features decrease during the
dip.  We have detected a systematic wavelength shift of spectral features
on the orbital period, but with a phase lag of $\simeq20\degr$, a
phenomenon that has been reported at optical wavelengths.  We discuss the
implications of our results in the context of models of an accretion disk
with a chromosphere between the disk and the extended wind. Finally, we
note that the observed FUV flux is too low to be consistent with the
temperature and radius of the WD derived by Baptista et al.  (1995),
suggesting that their remaining binary parameters, including a mass ratio
of 1, ought to be viewed with skepticism.


\end{abstract}

\section{Introduction}

Cataclysmic variables (CVs) are interacting binary systems in which a
late-type donor star overflows its Roche lobe and transfers mass to a
white dwarf (WD). In the non-magnetic CVs, mass is accreted through a
disk around the WD. Novalikes (NLs) are the non-eruptive class of CVs;
that is, they do not show dwarf novae outbursts.  They are believed to
have high mass-transfer rates, which keeps their disks in a permanent
outburst state.  The accretion disks in NLs are typically well
described by optically thick, steady-state disk models with accretion
rates $\dot{M} = 10^{-9} - 10^{-8} \msun$~yr$^{-1}$ (e.g., Horne 1993).
Because their underlying accretion behavior is fairly well understood,
novalike CVs are excellent laboratories to study the more complicated
processes associated with disk accretion.

Whether in the form of collimated jets or winds, outflows are closely
linked to the disk accretion process.  Wind signatures --- broad,
blueshifted absorption and/or P~Cygni profiles in resonance lines such as
\ion{C}{4} $\lambda\lambda$1548,1552~\AA, \ion{Si}{4}
$\lambda\lambda$1393,1402~\AA, and \ion{N}{5}
$\lambda\lambda$1232,1238~\AA\ --- are seen in most high accretion rate,
disk-fed CVs, including dwarf novae in outburst and novalikes (e.g., Heap
et al.\ 1978; Holm, Panek \& Schiffer 1982; C\'{o}rdova \& Mason
1982). Typical terminal velocities in the wind lines range from 3000 --
5000~km~s$^{-1}$, near the escape velocity of the primary
\citep{cordova1982}.  In eclipsing systems, the resonance lines --- seen in
emission --- are only partially eclipsed or completely unocculted,
indicating that the winds are vertically extended (e.g., Mauche et al.\
1994).

Much of the phenomenology of disk accretion and outflow in CVs has been
revealed by study of the novalike CV, UX~UMa. UX~UMa has an orbital period
of 4.72~hr and is eclipsing (it is, in fact, the brightest eclipsing NL).
The binary parameters for UX~UMa are typically taken from
\citet{baptista1995}, and are based on their identification of WD eclipse
features in the UV light curve, a standard mass-radius relation, and
previously published limits on the $K_{1}$ amplitude of the WD
\citep{shafter1984}.  \citet{baptista1995} obtained an inclination of
$71\arcdeg$ and, surprisingly, a mass ratio $q = M_{2}/M_{WD} = 1$.  They
also determined a WD flux and a temperature --- 52,000~K, or 70,000~K if
the lower half of the WD is obscured by the disk --- that would indicate a
strong WD contribution in the FUV.

\citet{knigge1997} modeled the profile changes in the \ion{C}{4}
$\lambda\lambda$1548,1552~\AA\ doublet of UX~UMa through eclipse.  They
showed that modeling the line with a biconical disk wind model provides a
good fit to the line profile in and out of eclipse.  The \ion{C}{4} line
has narrow absorption reversals, or dips, that disappear at
mid-eclipse. \citet{knigge1997} showed that the orbital behavior of the
dips could be explained by the presence of a slow, dense transition region
(between the disk and the fast, extended wind) that is occulted during
eclipse. This transition region, or disk chromosphere, may correspond to
the chaotic, slowly-outflowing wind regions that have appeared in some
radiative transfer models of CV disk winds \citep{proga1998,proga1999}.

The structure and behavior of winds and chromospheres in UX~UMa and other
high state CVs is closely tied to unresolved questions concerning the
structure of the underlying accretion disk.  \citet{knigge1998} modeled
optical and UV spectra of UX~UMa and found that steady-state accretion disk
models provided poor fits to the shape of the continuum at blue wavelengths
and at the Balmer jump (see also Baptista et al.\ 1998).  The discrepancy
between disk models and the observed spectrum is ameliorated if significant
amounts of optically thin continuum emission are present in the system.  An
accretion disk chromosphere is a plausible, though currently ad hoc, source
of such emission.  Other possibilities have been suggested to resolve the
mismatch between disk models and the observed spectrum in UX~UMa and other
NLs, such as a disruption of the inner accretion disk that removes the
bluest emitting regions \citep{long1994}.

To date, virtually all knowledge of the properties of the wind in
UX~UMa and its relationship with the accretion disk are based on
observation and modeling of the \ion{C}{4} doublet. This is not ideal,
since we do not know whether the behavior of \ion{C}{4} is
representative of other wind-dominated lines.  The strength and
profile of \ion{C}{4} has been observed to change over a single orbit
and over longer timescales \citep{mason1995}.  This variability may be
due to asymmetries in the wind which cause orbital variations and/or
to secular fluctuations in the mass loss rate; alternatively, it may
indicate that there are components of the \ion{C}{4} line that do not
originate in the wind at all.

There is a wealth of further information that can be gathered by extending
high spectral and time resolution observations of UX~UMa to shorter
wavelengths. The FUV is an excellent window in which to probe the structure
of disks and winds in CVs. For typical maximum temperatures in high state
accretion disks ($T \sim 30,000$~K), the disk SED peaks in the FUV, so the
FUV spectrum is a sensitive tracer of changes in the temperature and
structure of the innermost regions of the disk.  The FUV is also richly
populated with spectral lines of a variety of elements and ionization
states which, in principle, can be used to determine the thermal and
ionization structure of the wind.

Consequently, we have pursued FUV spectroscopy of UX~UMa with the Far
Ultraviolet Spectroscopic Explorer (FUSE), which covers 905 -- 1182~\AA\ at
spectral resolutions (through its large aperture) near 20,000
\citep{moos2000}. We have also examined archival HST UV spectroscopy to
obtain a broader picture of the UV spectrum in UX~UMa. In this manuscript,
we present observations (\S\ref{sec_obs}) and analysis
(\S\ref{sec_analysis}) of the FUV spectrum of UX~UMa and discuss what the
observations reveal about the structure of disks and outflows in disk
accreting systems (\S\ref{sec_discussion}).  We present our conclusions in
\S\ref{sec_conc}.

\section{Observations and Data Reduction} \label{sec_obs}

We observed UX~UMa with FUSE on 2001 March 23 -- 25.  A summary of the
observations is given in Table~\ref{tab_obs}.  The observations were spread
over six binary orbits and covered each orbital phase at least twice.  The
FUSE telescope collects data through four optical channels which cover the
wavelength range of the instrument (905 -- 1182~\AA) with some
redundancy. (For a summary of the FUSE optical design and its on-orbit
performance, see Sahnow et al.\ 2000 or the FUSE Observer's
Guide\footnote{http://fuse.pha.jhu.edu/support/guide/obsguide.html}.)  The
four channels were co-aligned during our observations, so no correction to
the relative wavelength solutions or fluxes of the channels was necessary.
All of the data were acquired through the LWRS aperture ($30\arcsec \times
30\arcsec$) in time-tag, or photon counting, mode, in which each photon
event is recorded to an accuracy of 1~s.

We re-reduced the data using the latest version of the CALFUSE pipeline
(V.\ 2.0.5), extracting spectra in 200~s intervals.  For the orbital period
of UX~UMa \citep{baptista1995}, this corresponds to an orbital phase
resolution of 0.012~cycles per spectrum.  We used the 200~s spectra as our
base set of data, which we binned up in time and/or spectral resolution for
specific tasks. When combining spectra from the four optical channels, we
specified a linear dispersion scale for the output spectrum and added each
input datum to the appropriate output bin, weighting the bin average by the
sensitivity of the input pixels. Regions of poor flux calibration, most
notably the ``worm'' in channel LiF1, were excluded.

We also created high time resolution light curves of the observations
using the raw time-tag photon event files.  To construct the light
curves, we defined rectangular apertures for each optical channel and
summed all of the events occurring in each channel each second.  In
defining the apertures, we excluded Ly$\alpha$ airglow and the
``worm''.  We also screened photon events using the default pulse
height limits adopted by the CALFUSE pipeline. The detector
background, summed over astigmatic spatial height and dispersion
direction (see the FUSE Observer's Guide), is
$\sim$0.5~counts/s/segment at night and $\sim$0.8~counts/s/segment
during the day.  The background is not negligible compared to the FUV
count rate of UX~UMa ($\simeq 100$~counts/s), particularly during
eclipse, so we subtracted background emission from the counts in each
1~s time interval.  The background was estimated by defining apertures
away from the target spectrum of the same size as the target apertures
and summing the counts therein.

\section{Analysis}  \label{sec_analysis}

\subsection{The Time-Averaged FUV Spectrum}

\begin{figure*}
\psfig{file=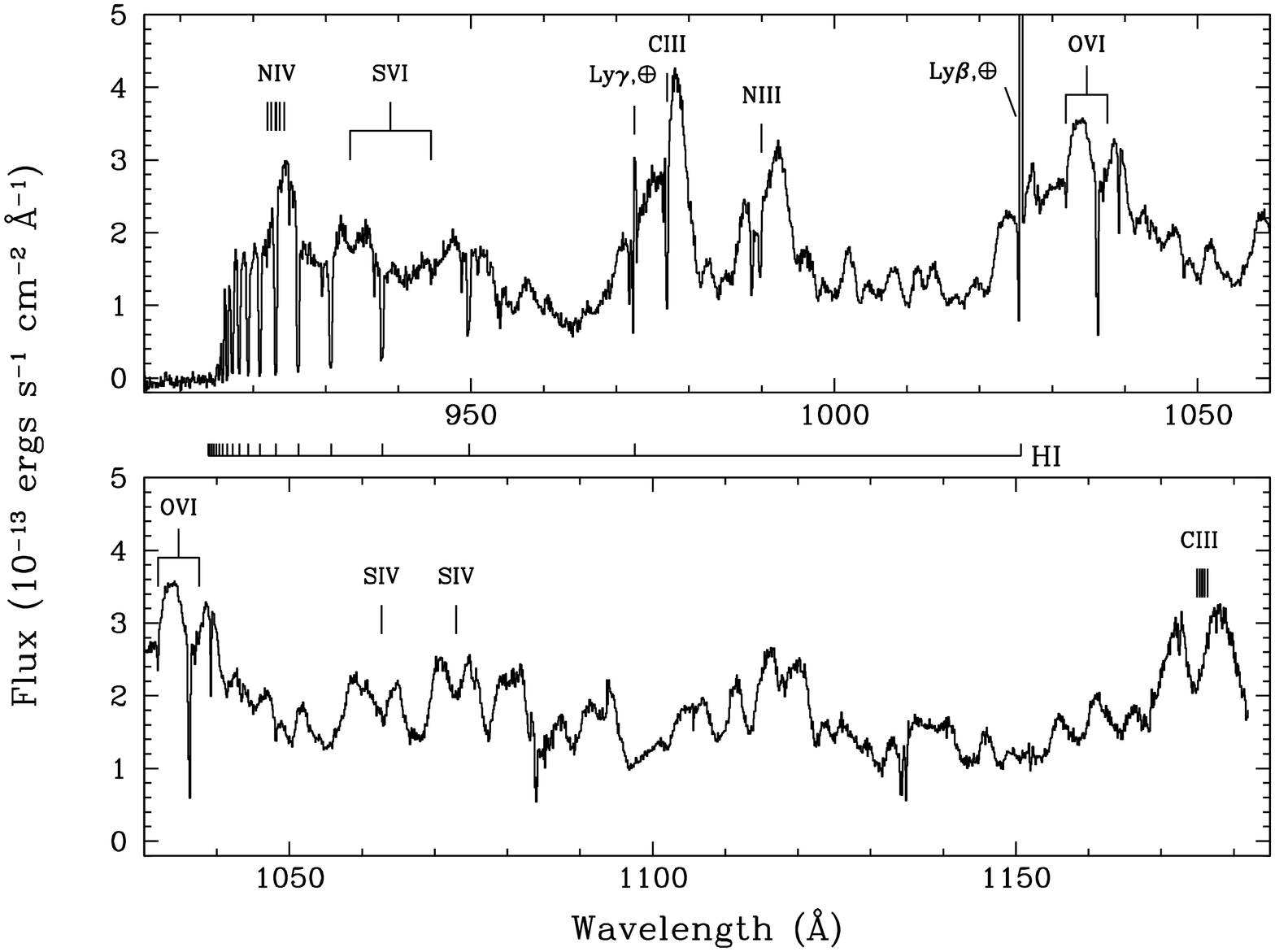,height=5in,angle=-90}
\figcaption[f1.eps]{The time-averaged FUV spectrum of UX~UMa from
2001 March.  The spectrum has been binned to 0.1~\AA\
dispersion. Prominent transitions from UX~UMa are marked and lines of
terrestrial airglow are labeled with circled crosses.
\label{fig_uxuma}}
\end{figure*}

The time-averaged spectrum of the complete data set is shown in
Figure~\ref{fig_uxuma}.  The spectrum has been binned to a dispersion of
0.1~\AA, which is the typical resultant resolution for a LWRS spectrum when
the channels are combined (see Froning et al.\ 2001). The spectrum is
dominated by strong, broad emission lines from a range of elements and
ionization levels. The strongest lines are resonance lines and
low-excitation energy excited state transitions of \ion{C}{3}
$\lambda$977~\AA\ and the \ion{C}{3} blend at $\lambda$1175~\AA; \ion{N}{3}
$\lambda$990~\AA; the \ion{N}{4} blend at $\lambda$923~\AA; and \ion{O}{6}
$\lambda\lambda$1032,1038~\AA. In between the strong emission lines, the
spectrum is populated by a host of weaker spectral features. The line
spectrum is quite complex, and a comparison of the time-averaged spectrum
with the spectrum at mid-eclipse (see Figure~\ref{fig_eclipse}) shows that
many lines overlap and are blended with each other.  Some of the other
lines that can be identified in the time-averaged and/or the mid-eclipse
spectrum are \ion{S}{6} $\lambda\lambda$933,944~\AA; \ion{P}{4}
$\lambda$951~\AA; \ion{S}{3} $\lambda$1013~\AA; \ion{S}{4} $\lambda$1063
and $\lambda$1073~\AA; \ion{He}{2} and/or \ion{N}{2} $\lambda$1085~\AA;
\ion{Si}{3} $\lambda\lambda$1108~\AA,1109~\AA, and 1113~\AA; and
\ion{Si}{4} $\lambda\lambda$1122,1128~\AA. There is a rise in flux at the
shortest wavelengths, $<$965~\AA.  Some lines that contribute to the
increase are identifiable, but they do not account for all of the emission
observed.  Excess emission near the Lyman limit is a common feature in the
FUV spectra of CVs: a similar bump is seen in the outburst spectrum of the
dwarf nova, WZ~Sge, and in the quiescent spectrum of SS~Cyg (Long et al.\
2002, 2002b, in preparation).  All of these spectra also show FUV lines in
emission, which suggests that the blue excess is a blend of line features,
dominated by emission from the upper order Lyman series lines.  Neutral
hydrogen emission is also evident in the presence of Ly$\beta$ and
Ly$\gamma$ emission features in the spectrum of UX~UMa.

The profiles of the strongest lines in the FUV spectrum of UX~UMa are
entangled with those of the weaker features, but a rough examination of
line shapes can be made.  The emission lines are broad, with FWHM $\geq
1800$~km~s$^{-1}$ (where the instrumental resolution is
$\simeq30$~km~s$^{-1}$). They have double-peaked profiles with a central
absorption reversal at the rest wavelength of each line. The line peaks are
separated by 1000 -- 1500~km~s$^{-1}$ and are roughly equidistant from the
line center.  In the strongest lines --- \ion{N}{4} $\lambda$923~\AA,
\ion{C}{3} $\lambda$977~\AA, \ion{N}{3} $\lambda$990~\AA, and \ion{O}{6}
$\lambda\lambda$1032,1038~\AA\ --- the red peak is enhanced relative to the
blue peak.  The FWHM of the absorption components of the lines are
$\simeq$600~km~s$^{-1}$. Both members of the \ion{O}{6} and \ion{S}{6}
doublets are present. We fit the Ly$\beta$ and \ion{O}{6} doublet blend
using the IRAF task ``specfit'' and assuming Gaussian profiles for all
three lines.  The equivalent widths of the two components of \ion{O}{6} are
equal, indicating that the lines are optically thick.  The FWHM of the
\ion{O}{6} lines (fixed to a 1:1 ratio) are 3800~km~s$^{-1}$.  In contrast
to \ion{O}{6}, the red component of the \ion{S}{6} doublet appears to be
weaker than the blue component, suggesting that these lines are not fully
optically thick.

Finally, the time-averaged spectrum also shows narrow interstellar
absorption lines from \ion{H}{1} and metal transitions. Identified
interstellar lines and their EWs are given in Table~\ref{tab_ism}.  We
pursued a curve of growth analysis to determine the interstellar
column density along the line to UX~UMa, resulting in an estimate of
the column density of $\log N{_H} = 20.3^{+0.5}_{-0.8}$.  The curve of
growth analysis is presented in Appendix~\ref{sec_cog}.

\subsection{Light Curves and Spectra in Eclipse} \label{sec_eclipse}

The out of eclipse and mid-eclipse spectra of UX~UMa are compared in
Figure~\ref{fig_eclipse}.  Also shown is the difference between the two
spectra; this is the spectrum of the eclipsed light.  The out of eclipse
spectrum was created by binning the time series 200~sec spectra into 80
phase bins (0.0125~cycles orbital phase resolution) and averaging the 18
bins preceding and following the eclipse (i.e., from orbital phases 0.85 --
0.95 and 0.05 -- 0.15).  It is qualitatively similar to, but slightly
brighter than, the time averaged spectrum; see \S~\ref{sec_lc} for a
discussion of orbital variations in the FUV spectrum away from the
eclipse. The mid-eclipse spectrum is the spectrum of the phase bin at the
light curve minimum.  The mid-eclipse spectrum and the difference spectrum
are binned to a resolution of 0.6~\AA. The eclipse spectrum is less complex
than the out of eclipse spectrum.  At mid-eclipse, the FUV spectrum is
characterized by broad emission lines separated by regions of continuum
emission. The average flux of the continuum regions of the mid-eclipse
spectrum is \expu{2.6}{-14}{$\flam$} and the continuum is relatively
constant over the FUV.

Many of the emission lines present in the out of eclipse spectrum remain
strong in eclipse, indicating that they arise from regions that are largely
unocculted.  The excess emission at wavelengths $<$955~\AA\ is not eclipsed
either. Many features that were not readily identifiable in the
time-averaged spectrum due to line blending can be identified at
mid-eclipse.  Mid-eclipse emission lines include transitions of \ion{C}{3},
\ion{N}{3}, \ion{N}{4}, \ion{O}{6}, \ion{Si}{3}, \ion{Si}{4}, \ion{S}{3},
and \ion{S}{4}.  There is an emission line at $\lambda$1085~\AA\ from
\ion{He}{2} and/or \ion{N}{2} and an unidentified feature at
1143~\AA. Another emission feature at $\lambda$951~\AA\ may be from
\ion{P}{4}, although \ion{H}{1} $\lambda$949~\AA\ is also present in that
region.  Because of the presence of clean, line-free regions in the eclipse
spectrum, we were able to measure the parameters of most of the emission
lines seen at mid-eclipse. The EWs and line widths of these lines are given
in Table~\ref{tab_eclipse}.

A comparison of the out of eclipse and mid-eclipse spectra indicates that
most of the weaker, blended features are occulted in eclipse. The
difference spectrum shows that the lines are a complex blend of components,
predominately in absorption. The line absorption can be just as strong or
stronger than the line emission, so that some lines, like the \ion{Si}{4}
doublet, can only be identified at mid-eclipse, after their absorption
components are occulted. Along with the weaker features, the absorption
reversals seen in the profiles of strong lines in the time-averaged
spectrum are eclipsed.

We searched for evidence of a rotational component in the lines by
examining changes in the FUV spectrum through the eclipse and comparing the
profiles of the strong emission lines in the out-of-eclipse and mid-eclipse
spectra. If the line regions are rotating in Keplerian motion, the blue
wing and peak should be occulted earlier than the red peak and wing. This
was seen in \ion{C}{4} $\lambda\lambda$1548,1552~\AA\ in UX~UMa
\citep{baptista1995,knigge1997}.  Most of the lines in the FUV spectrum are
too confused with other features to examine their eclipse behavior in
different velocity bins, but we examined the eclipse of two relatively
unblended lines, \ion{C}{3} $\lambda$1175~\AA\ and \ion{S}{4}
$\lambda$1073~\AA.  In both of these lines, the FUV eclipse minimum occurs
at earlier phases at blue velocities relative to the red components of the
line.  There also appears to be some narrowing of the lines in eclipse: the
FWHM of \ion{C}{3} and \ion{S}{4} $\lambda$1073~\AA\ decreases from
$\sim$3000~km~s$^{-1}$ to $\sim$2000~km~s$^{-1}$ during the eclipse. A
narrower line profile during eclipse has been seen in \ion{C}{4} in UX~UMa
\citep{mason1995,knigge1997}.  Kinematic models predict narrowing of line
profiles in eclipse for a collimated disk wind viewed at high inclination,
as the more rapidly-rotating parts of the wind near the disk plane are
occulted (Knigge \& Drew 1997 and sources therein).

\begin{figure*}
\plotone{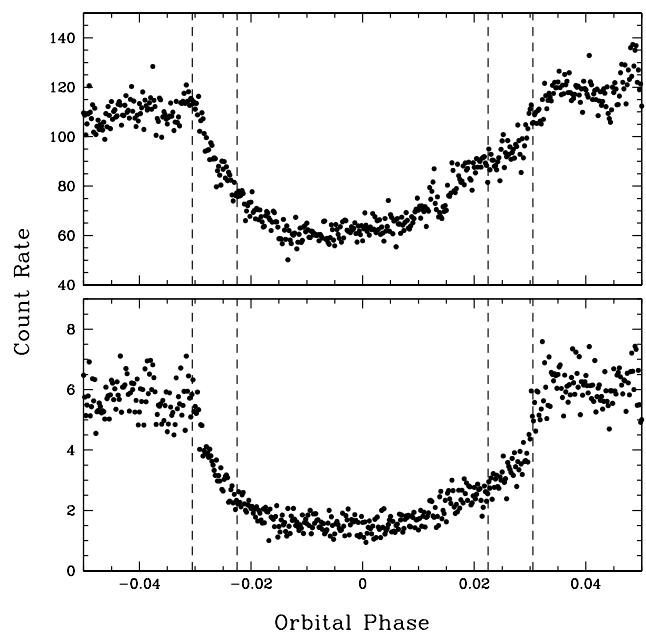}
\figcaption[f3.eps]{High time resolution light curves of the eclipse
in UX~UMa.  The upper panel shows the mean count rate over full 910 --
1182~\AA\ range through eclipse.  The lower panel shows the mean count
rate in four regions away from the strong emission lines and dominated
by continuum emission at mid-eclipse: 955 -- 965~\AA, 1045 --
1057~\AA, 1090 -- 1103~\AA, and 1149 -- 1164~\AA.  Both light curves
have been made by phase-binning the 1~s count rate light curves into
5000 phase bins per orbit (0.0002~cycles or 3.4~s per bin).  The light
curves have been shifted by 0.006 cycles relative to the published
photometric ephemeris to bring the phases of full width at half flux
in the lower, continuum-dominated light curve, symmetric about orbital
phase 0. Also shown for reference are the contact phases of the WD
eclipse found by Baptista et al.\ 1995. \label{fig_eclc}}
\end{figure*}

Figure~\ref{fig_eclc} shows phase-binned FUV light curves of the eclipse in
UX~UMa.  The upper panel shows the light curve over the full 910 --
1182~\AA\ range, while the lower panel shows the mean light curve for four
regions of the spectrum in which only continuum emission is present at
mid-eclipse: 955 -- 965~\AA, 1045 -- 1057~\AA, 1090 -- 1103~\AA, and 1149
-- 1164~\AA. We created both light curves by phase-binning the 1~s count
rate light curves into 5000 bins per orbit. We shifted both light curves
earlier in phase by 0.006~cycles to bring the phases of full width at half
depth of the lower, continuum-dominated, light curve symmetric about phase
0. This phase shift is typical of small-scale variations in the phasing of
the eclipse as indicated by the scatter in the (O--C) diagram for the
ephemeris \citep{baptista1995}.  Also shown in Figure~\ref{fig_eclc} are
the contact points for the eclipse of the WD as found by
\citet{baptista1995}.

Figure~\ref{fig_eclc} indicates a narrow eclipse in the FUV.  The full
eclipse width from start to end is $\simeq$0.065 cycles in both light
curves.  The full width at half depth for the upper light curve is 0.043
cycles, but is 0.055 cycles for the lower light curve, suggesting that the
emission lines are slightly more centrally concentrated than the continuum
emission. The upper light curve is also more asymmetric than the lower
curve: the minimum occurs at earlier phases in the former, and its egress
is longer than its ingress. Both eclipses are fairly smooth, although there
is a hint of a break around phase 0.0175 that could be an egress feature.
The FUV eclipse is not total, although it is much deeper in the regions
between the resonance emission lines.  The eclipse depth over the full FUV
range is 40\% of the pre-eclipse flux, but it increases to an average depth
of 82\% for the four regions dominated by continuum emission.\footnote{The
eclipse depths quoted were taken from flux-calibrated light curves of the
eclipse.  The raw count rate light curves shown in Figure~\ref{fig_eclc}
give slightly shallower eclipses of 45\% and 73\% for the upper and lower
curves, respectively.  The discrepancy between fluxed and raw eclipse
depths most likely indicates that our method for estimating the background
when creating the high time resolution light curves underestimated the true
background.}  The residual continuum emission at mid-eclipse probably
originates from the back of the accretion disk, which is not occulted
during eclipse. Emission from disk annuli as close as 11--14~R$_{WD}$ from
the center of the primary is still visible at mid-eclipse (the lower value
assumes the system parameters of Baptista 1995, while the upper value is
for the geometry of Smak 1994, and using M$_{WD}/R_{WD}$ from Nauenberg
1972).

\subsection{Orbital Variability Outside of the Eclipse} \label{sec_lc}

Past analysis of time and orbital variability outside of the
eclipse has shed light on behavior of the inner disk and outflow
in UX~UMa. For example, UX UMa is known to occasionally exhibit
low-amplitude 28-s oscillations in optical and UV light (e.g.
Warner \& Nather 1972; Nather \& Robinson 1974; Knigge et al.\
1998b). The oscillations originate very near or on the WD surface,
and may indicate an appreciable magnetic field that affects the
structure of the inner disk.  On longer time scales,
\citet{mason1997} found an orbital variability in the UV light
curve of UX~UMa in the form of a decline in flux, or ``light curve
dip'', centered on orbital phase 0.6. They attributed the
phenomenon to asymmetric vertical extension in the outer accretion
disk absorbing flux from the disk interior. They found a decrease
in absorption line EWs during the dip, which indicates that the
narrow absorption components seen in the UV lines in UX~UMa occur
not in the outermost accretion disk annuli but interior to the
source of the orbital light curve dip.

\begin{figure*}
\psfig{file=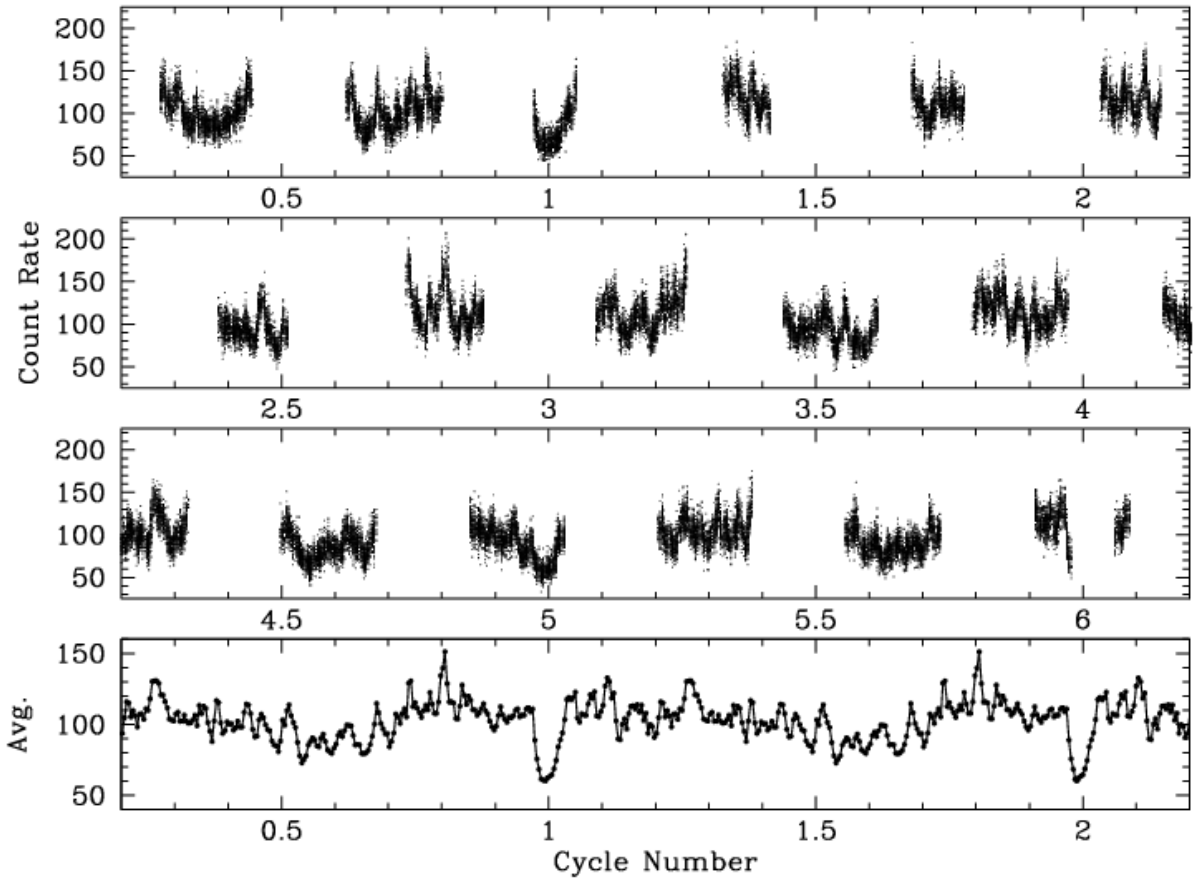,height=5in,angle=-90}
\figcaption[f4.eps]{High time resolution light curve of the 2001
March FUV observations of UX~UMa and a phase-binned light curve of the
binary orbit.  The top three panels show the count rate in 1~s bins
for the FUSE observations of UX~UMa.  The count rate is shown versus
cycle number of the observation.  The bottom panel shows the count
rate data binned on the orbital period into 250 phase bins.  The
binned light curve is repeated twice. \label{fig_contlc}}
\end{figure*}

\begin{figure*}
\psfig{file=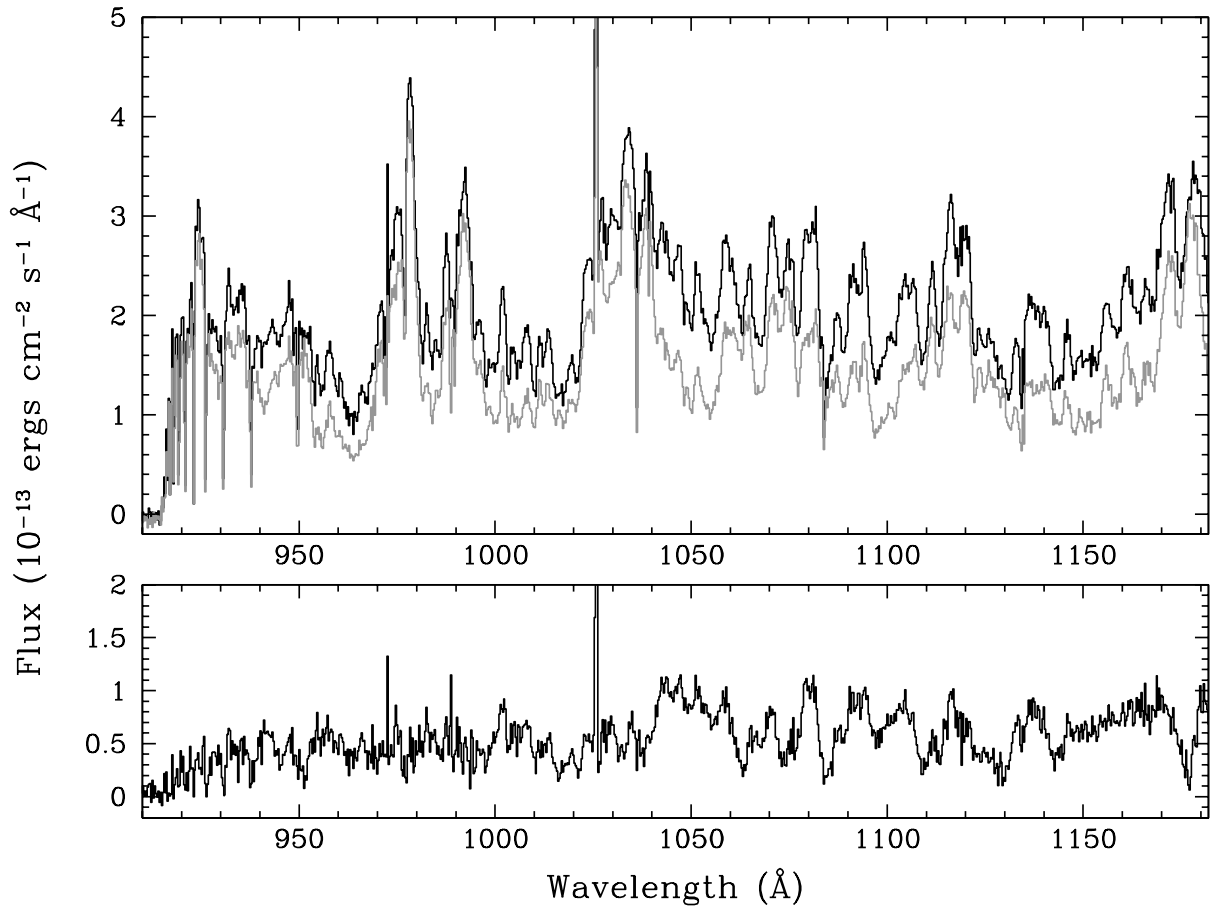,height=5in,angle=-90}
\figcaption[f5.eps]{FUV spectral variations during flickering
events.  The upper panel shows in black the average of all the 200~sec
spectra acquired at or near the peak of flickers.  The spectrum shown
in gray is the average of the 200~sec spectra acquired in low count
rate intervals, away from flickers.  Spectra acquired in eclipse were
excluded from both averages.  The bottom panel shows the difference
spectrum between the flicker and non-flicker
spectra. \label{fig_flare}}
\end{figure*}

In an effort to search for the oscillations and orbital variability in
UX~UMa in the FUV, we examined the high time resolution light curves
created from the raw photon event files of the FUSE observations and
spectral variations in the 200~sec spectra.  The top three panels of
Figure~\ref{fig_contlc} show the FUV light curve of UX~UMa in 2001
March, with the count rate in each 1~s bin plotted vs.\ cycle
number. The bottom panel shows the same data binned into 250 phase
bins per orbit. The 1~s light curve shows significant flickering in
the FUV light on short time scales, with flickering peaks typically
separated by 200 -- 1000~s.  In several epochs, the count rate changed
by a factor of 2 or more within a 5~min time period
($\simeq$0.02~cycles).  We compared the spectrum of the flickers to
the spectrum away from the flickers by averaging all of the 200~sec
spectra that occurred at or near the peak of a flicker and all of the
200~sec spectra obtained between flickers (excluding the eclipse). The
``flicker'' and ``non-flicker'' spectra are shown in
Figure~\ref{fig_flare}.  Also shown is the difference spectrum between
the high and low states.  The difference spectrum, which is the
spectrum of the flickering light, is quite similar in morphology to
the spectrum of the eclipsed light at mid-eclipse.  That is to say,
the source of the flickering is the same material, located near the
accretion disk plane, that is occulted in eclipse.  Although the
continuum shows strong flickering, the prominent emission lines do not
flicker.

To search for periodicity in the FUV flux variations, we carried
out a power spectrum analysis of the 1~s overall FUV light curve.
No sign of the 29~s oscillations was found.  We also failed to
detect an unambiguous periodicity to the flickering. By injecting
sinusoidal signals of varying amplitude into the data stream, we
found that we would certainly (probably) have detected the 29~s
oscillations with an amplitude of 3\% (1\%) in our data set. For
comparison, the average amplitude of the oscillations in Knigge et
al.'s (1998b) HST data (which covered 2000\AA\ -- 8000 \AA) was
roughly 0.5\%. However, in their data set, the oscillation
amplitude rose to well over 1\% near 2000~\AA, due to the
extremely blue spectrum of the oscillations. It therefore seems
likely that we would have been able to detect oscillations of the
same strength and spectral shape as those seen by Knigge et al.
The absence of the oscillations from our data is not particular
surprising, however. For example, Knigge et al.\ saw the
oscillations in only two of their four observing runs.

The phase-binned FUV light curve in Figure~\ref{fig_contlc} shows
that in addition to the short time scale flickering, there is an
orbital modulation in the form of a decline in flux centered on
phase 0.55. On average, the FUV flux is 18\% lower between phases
0.45 -- 0.65 compared to the rest of the orbit outside of eclipse,
and it reaches a maximum depth of 25\%.  The depth of the eclipse
over the full FUV is 40\%, so the light curve dip is over half the
eclipse depth in the FUV.  The dip in the FUV light curve occurs
at roughly the same orbital phases as the dip observed by
\citet{mason1997} in 1996 November HST/GHRS spectra of UX~UMa
(these spectra are discussed further in \S\ref{sec_hst}).  The
1996 observations did not have full coverage of the orbit, but on
two epochs (different binary orbits) the flux between phases 0.6
-- 0.7 was 45\% and 60\% lower than the flux at orbital phases
0.25 -- 0.4 (measured in the 1415 -- 1433~\AA\ region), a deeper
modulation than seen in the FUV.

\citet{mason1997} also noted that the EWs of the absorption
features in \ion{Si}{4} $\lambda\lambda$1393,1402~\AA\ decreased
during the phases of the dip.  We compared the FUV spectrum in the
dip phases (0.45 -- 0.65) to the spectrum outside of the dip (0.05
-- 0.25 and 0.75 -- 0.95). As with the short time scale
flickering, the spectral changes in the orbital dip are seen in
the weak lines and the continuum and not in the prominent line
emission. The EWs of the absorption components of the FUV lines
decrease during the light curve dip, consistent with the results
of Mason et al. that place the material responsible for the line
absorption interior to the source of the broadband light curve
dip.  This conclusion is based on UV and FUV observations that
only covered the binary orbit approximately twice each, however,
and an observation covering multiple binary orbits is necessary to
ensure that orbital effects are not contaminated by the spectral
variations from flickering.

\begin{figure*}
\plotone{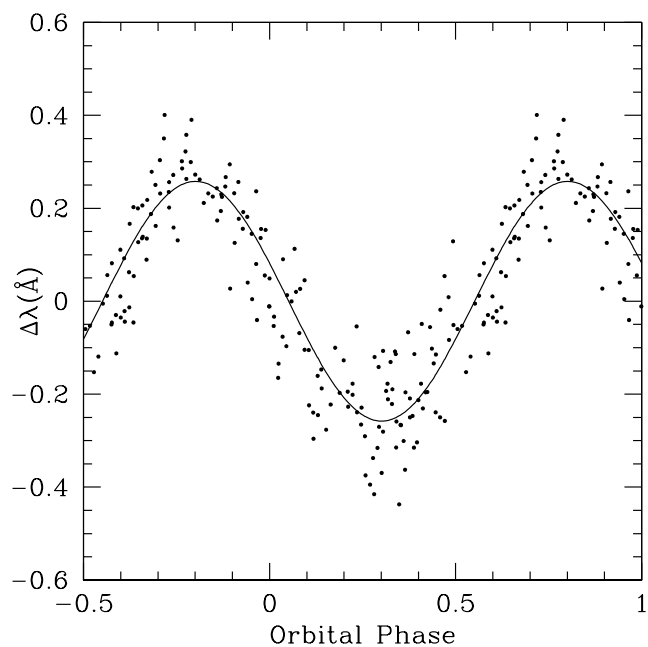}
\figcaption[f6.eps]{Radial velocity variations over the binary
orbit in UX~UMa.  The points plotted show the wavelength shifts
for each of the 200~s spectra from the 2001 March observation
relative to the time-averaged spectrum of the observation.  The
solid line is the best-fit sinusoid.  It has an amplitude of
0.26~\AA\ and a phase shift of 0.057~cycles. \label{fig_vrad}}
\end{figure*}

Finally, we searched for orbital radial velocity variations in the FUV
spectra using the 200~s spectra.  With the ``crosscor'' task in IRAF, we
cross-correlated the individual spectra with respect to the time-averaged
spectrum of the observation in several different wavelength regions.  We
also measured the centroid of several of the narrow absorption features in
the lines by fitting a Gaussian to their profiles in each spectrum.  A
representative example of the results is shown in Figure~\ref{fig_vrad},
which plots the wavelength shift of each 200~s time series spectrum as a
function of orbital phase (corrected by $-$0.006~cycles; see
\S\ref{sec_eclipse}) for the 1040 -- 1182~\AA\ region (this region is shown
because this part of the spectrum is largely free of non-varying features
from interstellar lines and airglow).  The best-fit sinusoid, also shown in
Figure~\ref{fig_vrad}, has an amplitude of 0.26~\AA\ (or a total
peak-to-peak shift of 0.52~\AA) and a phase shift of 0.057~cycles. The
amplitude corresponds to a velocity of 70~km~s$^{-1}$.  The phase shift
gives a 21$\arcdeg$ lag between the photometric phasing and crossing times
in the line spectrum.  In general, this amplitude and phase shift are
descriptive of the orbital radial velocity variations of the FUV spectrum.
One possible exception are the absorption reversals in the lines, which
show much larger amplitudes, from 140~km~s$^{-1}$ for the \ion{S}{4} and
\ion{Si}{3} absorption features to a peak amplitude of 200~km~s$^{-1}$ for
the absorption component of \ion{C}{3}. The wavelength shift in the
absorption lines shows the same phase offset as seen in the broader
spectrum.  We caution that, due to the complex line blending in the FUV
spectrum, the apparent change in the velocities of the absorption component
centroids could be an artifact of a profile change in the emission
components of the lines.

The amplitude of the orbital variation in the overall FUV spectrum is
smaller than the value of $K_{1} \geq 150$~km~s$^{-1}$ found by
\citet{shafter1984} from radial velocity variations in the emission wings
of H$\alpha$ (although the FUV absorption lines may show a range of motion
equal to or larger than the $K_{1}$ determination). \citet{schlegel1983}
performed a spectrophotometric study of the optical lines in UX~UMa,
observing it twice separated by a year.  Although they did not observe a
velocity amplitude as low as 70~km~s$^{-1}$, they found that the optical
lines exhibited a range of radial velocity amplitudes (121 --
267~km~s$^{-1}$) that varied from line to line and within components of the
same line from one epoch to another.  More interesting, Schlegel et al.\
observed a 0.06~cycle phase shift, consistent with the lag we observe in
the FUV, in the radial velocity curve of \ion{He}{2} $\lambda$4686~\AA\ in
their 1981 observations.  The phase lag was not present in 1982. Because of
the presence of a pre-eclipse light curve hump in 1981 but not 1982,
Schlegel et al.\ attributed the phase lag at the earlier epoch to a
substantial bright spot component in \ion{He}{2}. There were problems with
this interpretation, however, namely that there was no S-wave component and
no eclipse in the \ion{He}{2} line in 1981, both of which are expected for
bright spot emission. As a result, they suggested that \ion{He}{2} emission
region may have had substantial vertical extent in that epoch. The phase
shift in the FUV between the wavelength shifts and the photometric
ephemeris likely derives from some asymmetry in the line formation region,
but while the phasing is suggestive of an asymmetry tied to the bright
spot, the FUV emission originates interior to the outside edge of the disk,
so it is not obvious how to reproduce the observed amplitude and phase
shifts.

\subsection{Archival HST Spectra of UX~UMa} \label{sec_hst}

As noted earlier, UX~UMa has previously been observed at UV
wavelengths with HST.  Of particular interest for its combination
of moderate spectral resolution and broad wavelength coverage is
the 1996 November observation acquired with the GHRS.  This
observation obtained 1149 -- 1660~\AA\ wavelength coverage at
0.6~\AA\ spectral resolution.  The spectra were obtained in RAPID
mode with 5~s integration time and sampled much of the binary
orbit. Orbital light curves and spectra of \ion{Si}{4}
$\lambda\lambda$1393,1402~\AA\ from this data set were presented
by \citet{mason1997} as part of an analysis of the pre-eclipse
dips discussed in \S~\ref{sec_lc}, but the full UV spectrum was
not published.  As these spectra provide a longer-wavelength
complement to our FUV observations, we present them here.

Figure~\ref{fig_eclipse_hst} shows the time-averaged, mid-eclipse,
and difference spectra of the 1996 GHRS observation of UX~UMa. The
mid-eclipse spectrum covers the central 0.0125~cycles of the
eclipse. The HST time-averaged spectrum is similar in character to
the FUSE FUV spectrum: in both, continuum emission is everywhere
overlaid by a rich spectrum of lines with emission and absorption
components.  Strong resonance lines of \ion{C}{4}, \ion{N}{5}, and
\ion{Si}{4}, as well as \ion{He}{2}, are joined by a wealth of
other lines, including transitions of \ion{C}{3}, \ion{Si}{2} and
\ion{Si}{3}.  The strong emission lines have line profiles similar
to those seen in the FUV. The lines are double-peaked, with
central absorption at the rest wavelengths of the lines.  At this
epoch, the absorption reversal is relatively shallow in \ion{C}{4}
$\lambda\lambda$1548,1552~\AA, \ion{N}{5}
$\lambda\lambda$1238,1242~\AA, and \ion{He}{2} $\lambda$1640~\AA.
The absorption dips are deeper in \ion{C}{3} $\lambda$1175~\AA,
\ion{C}{2} $\lambda$1334~\AA, and \ion{Si}{4}
$\lambda\lambda$1393,1402~\AA. In \ion{Si}{2} and \ion{Si}{3}, the
lines are primarily or entirely in absorption outside of eclipse.

The FUSE and GHRS spectra overlap from 1149 -- 1182~\AA, a region
including the \ion{C}{3} $\lambda$1175~\AA\ blend.  A comparison
of the line at two epochs indicates that the profile of the line
is variable.  Although the line is double-peaked in both spectra,
the blue peak dominates in the 1996 spectrum, while the red peak
is stronger in 2001.  Spectral minima (around 1150~\AA, for
example) are at the same flux in both spectra, as is the blue peak
in \ion{C}{3}, but other spectral peaks, at 1158~\AA, 1162~\AA,
and the red peak of \ion{C}{3}, are depressed by 15\% -- 25\% in
the 1996 observation relative to 2001.  The \ion{C}{4} line also
shows a strong peak blueward of the rest wavelengths of the
doublet lines. This enhanced blue peak is not present in other
high-resolution spectra of \ion{C}{4} (1993 GHRS spectra, see
Mason et al.\ 1995, and unpublished 1998 STIS spectra).

The mid-eclipse HST spectrum is also quite similar to the FUSE FUV eclipse
spectrum.  The weaker line emission is gone, and the spectrum consists of
strong emission lines separated by a flattish continuum. The mean continuum
flux in the overlap region, 1149 -- 1165~\AA, is \expu{4.8}{-14}{$\flam$}
in the HST spectrum, compared to \expu{2.7}{-14}{$\flam$} in the FUSE
observation.  The UV flux (and, by implication, the mass accretion rate)
fluctuates in UX~UMa (see, e.g., Knigge et al. 1998) and the difference in
continuum flux between the 1996 and 2001 observations is typical for the
system. In \ion{C}{3} $\lambda$1175~\AA, the line profile is reversed in
the HST spectrum compared to the profile in the FUSE observation, with the
blue peak dominant in the former and the red peak in the latter. The
difference spectrum in the HST observation shows that the eclipsed
component of the lines, as in the FUSE observations, are mainly or entirely
in absorption. (Note that here, because of phase coverage gaps, the
difference spectrum is the difference between the time-averaged spectrum
and the mid-eclipse spectrum, rather than a direct comparison of
mid-eclipse to phases immediately preceding and following.) The absorption
components are narrow, with typical FWHM of 600--900~km~s$^{-1}$, and very
deep: several of the dips reach zero flux at their line centers, which did
not occur in the FUSE observations. The absorption dips are occulted in
some lines, such as \ion{C}{3}, \ion{Si}{4}, and \ion{C}{4}, but remain at
diminished EW in other lines, including \ion{N}{5} and \ion{Si}{3}. The
broad \ion{Si}{3} blend at $\lambda$1300~\AA\ appears weakly in emission
only after its broad, deep absorption component is eclipsed.

\begin{figure*}
\plotone{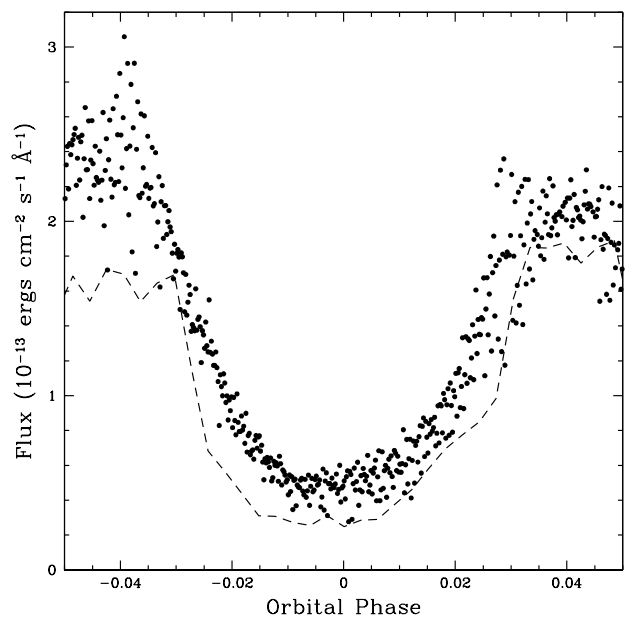}
\figcaption[f8.eps]{Eclipse light curve for the 1996 November HST
observations of UX~UMa.  The data points show the mean flux in four
regions away from the emission lines and dominated by continuum
emission at mid-eclipse: 1149 -- 1164~\AA, 1273 -- 1288~\AA, 1440 --
1493~\AA, and 1565 -- 1596~\AA.  The light curves was made by
phase-binning the fluxed spectra into 5000 phase bins per orbit
(0.0002~cycles or 3.4~s per bin).  The light curves have been shifted
earlier in phase by 0.0095 cycles relative to the published
photometric ephemeris to bring the phases of full width at half flux
of the light curve symmetric about orbital phase 0 (the out of eclipse
flux was assumed to be the mean of the pre- and post-eclipse flux
levels).  Shown for reference with the dashed line is the FUV eclipse
light curve from the 2001 FUSE observations.  The FUV light curve is
the mean flux from the same wavelength regions used for the lower
panel of Figure~\ref{fig_eclc}, but created from the flux-calibrated
spectra, extracted at 50~sec intervals through the eclipse, rather
than the raw count rate files (consequently, it's time resolution is
more coarse than the light curve in Figure~\ref{fig_eclc}).
\label{fig_eclc_hst}}
\end{figure*}

Figure~\ref{fig_eclc_hst} shows the light curve of the HST
observations for four regions of the spectrum away from the prominent
lines and dominated by continuum emission at mid-eclipse.  Also shown
for reference is the fluxed eclipse light curve from the 2001 FUSE
observation (at lower time resolution than the light curves in
Figure~\ref{fig_eclc} but created from flux-calibrated spectra).  The
pre-eclipse flux was higher than the post-eclipse flux in 1996, while
the opposite trend was true in 2001.  In other respects, the UV and
FUV light curves are quite similar, however.  The full width at half
depth of the 1996 UV eclipse (where the out of eclipse flux was taken
as the mean of the pre- and post-eclipse fluxes) is 0.052~cycles,
consistent with the eclipse width in the FUV. Moreover, the depth of
the eclipse does not change significantly with wavelength: the mean
depth in the continuum-dominated regions is 82\% in the FUV and 77\%
in the UV.  Even in the longest wavelength range, 1565 -- 1665~\AA,
the depth of the eclipse remains 70\% of the out of eclipse flux.

\section{Discussion} \label{sec_discussion}

The FUV spectrum of UX~UMa is everywhere characterized by line
features.  The FUSE and archival HST spectra show that from 910 --
1600~\AA, the time-averaged spectrum is composed of prominent emission
lines and a complex blend of weaker features. The emission lines arise
from the usual transitions seen in CV spectra in the UV: resonance
lines of neutral hydrogen and ionized metals and prominent excited
state transitions. In eclipse, these lines are largely unocculted,
while the weaker features are completely eclipsed. The spectrum of the
eclipsed light suggests that the weaker line features are primarily in
absorption.

Because of the severe line blending, it is difficult to determine
where to place the continuum in the FUV and UV spectra, and some of
the eclipsed line region may be in emission as well as in absorption.
At mid-eclipse, the spectrum in between the prominent resonance
emission lines is smooth and appears to consist only of continuum
emission.  As a result, the emission line component of the eclipsed
light, if present, must have substantial asymmetry or originate very
close to the WD to explain why it is not seen during eclipse, when
much of the back of the accretion disk remains unocculted. A possible
source for asymmetric emission is the mass accretion stream.
\citet{baptista1998} created spatially resolved spectra of the
accretion disk in UX~UMa from UV and optical HST/FOS observations.
They found that the quadrant of the disk containing the mass accretion
stream is a source of UV emission, including \ion{C}{4} line emission,
down to 0.1R$_{L_{1}}$, which they attributed to overflow of the mass
stream above the disk.  The introduction of an asymmetry in the line
emission by the mass stream is also consistent with the phase lag in
the orbital radial velocity variations in the FUV spectrum with
respect to the photometric ephemeris.

The origin of the line absorption also has important implications for
the structure of the accretion disk and the wind.  \citet{knigge1997}
modeled the \ion{C}{4} $\lambda\lambda$1548,1552~\AA\ line profile in
UX~UMa in and out of eclipse with a kinematic model of a rotating,
biconical disk wind.  They found that to match the eclipse of the
narrow absorption features at the rest velocity of the doublet, their
model had to include a dense, slowly outflowing transition region
between the fast wind and the disk, which they dubbed an accretion
disk chromosphere. Some hydrodynamical models of radiatively-driven
disk winds have predicted regions of chaotic, low velocity motion that
may be the source of the chromospheric features seen in UX~UMa
\citep{proga1998}.  The results of Knigge \& Drew were based on only
one epoch of observation of UX~UMa and only one line, however, which
left open the possibility that secular variability and/or non-wind
contributions to \ion{C}{4} could be responsible for the narrow
absorption components of the line.  An alternate explanation for the
absorption components is absorption by low-velocity material in the
outer disk; narrow, low-velocity absorption lines from the outer disk
are present in the FUV spectrum of the dwarf nova U~Gem in outburst,
for example \citep{froning2001}.

\citet{mason1997} showed that the narrow line absorption in UX~UMa
also occurred in \ion{Si}{4} $\lambda\lambda$1393,1402~\AA\ and that
the absorption line EWs decreased during a broadband light curve dip
that they attributed to asymmetric vertical extension in the outer
disk. The EW decline suggested that the line absorption components
were interior to the outer edge of the disk, which made the disk
wind/chromosphere model of \citet{knigge1997} more attractive. The
FUSE observations, obtained 5 years after the HST spectra, show that
the line absorption components are present throughout the UV and FUV
in a wide variety of species and ionization states and are persistent.
The EWs of the absorption lines in the FUV spectrum also decrease
during the broadband light curve dip. If the line absorption occurred
in the outer disk edge, the line EWs would increase, not decrease,
when the amount of absorbing material increased around phase 0.6.  The
low velocities and narrow profiles of the absorption lines are
consistent with origin in the dense, slowly outflowing transition
region between the accretion disk and the fast wind.  Nevertheless,
both the HST and FUSE observations sample a small number of orbits, so
it remains possible that the orbital spectral changes are being
confused with variability caused by flickering; additional broadband
UV observations spread over the binary orbit of UX~UMa will be
necessary to properly average out the flickering effects.

Another ongoing question in UX~UMa and in other CVs concerns the
continuum source in the ultraviolet.  In UX~UMa, as in other NLs, the
accretion disk is believed to be in a high mass-accretion rate
steady-state, and we expect it to be the dominant continuum source in
the optical and UV.  Previous studies have not been successful in
modeling the continuum emission in UX~UMa with steady-state disk
models, however. \citet{knigge1998} explored model disk fits to
broadband HST spectra (1250 -- 9000~\AA) of UX~UMa.  Their models were
constructed from sums of appropriately weighted stellar atmosphere
spectra at the temperature and gravity of each accretion disk annulus.
They demonstrated that the model disk spectra provide a poor fit to
the overall spectral shape.  At optical wavelengths, the models overestimate
the amplitude of the Balmer jump, while in the UV, the models are too
blue to match the observed spectrum.  The discrepancy at blue
wavelengths between model and observed spectra becomes worse when the
FUV spectrum is also taken into account: the model disk spectrum of
\citet{knigge1998} would exceed the observed 2001 March FUV flux by
$>$50\%.

We examined the FUV fluxes predicted by accretion disk models for the
geometry of UX~UMa (testing the geometries of both Baptista et al.\ 1995
and Smak 1994).  Optical and UV eclipse mapping analyses have given mass
accretion rates for UX~UMa from \expu{5}{-9}{\msun \: yr^{-1}} to
\expu{1}{-8}{\msun \: yr^{-1}}
\citep{rutten1992,baptista1995,baptista1998}, where the range reflects a
real variation in the observed flux and a likely variation in the mass flow
rate through the disk.  The distance to UX~UMa has been estimated to lie
from 215 -- 345~pc (Baptista et al.\ 1995 and sources therein), and the
reddening is E(B--V) = 0, or at most E(B--V) $<$ 0.04 \citep{knigge1998}.
We compared the fluxes of FUV and UV spectra of UX~UMa to model disk
spectra constructed from summed, area-weighted, Doppler-broadened spectra
of stellar atmospheres set to the appropriate temperature and gravity for a
given disk annulus, assuming the standard, steady-state temperature
distribution in the disk (see Froning et al.\ 2001 for a full description
of the models). 

There is a continued discrepancy between the models and the observed fluxes
at blue wavelengths.  If the accretion disk is assumed to be the dominant
FUV continuum source --- which corresponds to scaling the model to
something like the minima in the FUV spectrum away from the strong lines:
\expu{8}{-14}{$\flam$} at 965~\AA\ and \expu{1}{-13}{$\flam$} at 1095~\AA\
--- the predicted mass accretion rate is \expu{1.6}{-9}{\msun \: yr^{-1}},
assuming the upper range to the distance (increasing the reddening to the
upper limit gives $\dot{m} =$ \expu{2.2}{-9}{\msun \: yr^{-1}}).  Although
the 1996 November and 2001 March HST and FUSE spectra have the same out of
eclipse fluxes in their region of overlap (1149 -- 1182~\AA), the same
accretion disk model will not fit both spectra. The $\dot{m} =$
\expu{1.6}{-9}{\msun \: yr^{-1}} model that scales to the flux in the FUV
only provides half of the flux observed in 1996 November at 1250~\AA\ and
one third of the observed flux at 1400~\AA.  Increasing the reddening to
E(B--V) = 0.04 has a negligible effect on the slope of the spectrum in the
UV.  The unknown degree of line blanketing affecting the FUV and UV limits
our ability to assess the discrepancy between the observed continuum and
the disk models, but it is clear that a single accretion disk model cannot
fit the continuum over the full FUV+UV range.

Although the discrepancy between the disk models and observed spectra in
high state CVs could result from some problem with the physics of the
models, a more likely cause is an incomplete description of the
three-dimensional structure of the accretion disk.  The presence of
substantial vertical extent in the disk, as indicated by the absorption
components in the spectral lines and the orbital dip at phase 0.6, points
to a more complex accretion disk structure than current disk models assume.
\citet{knigge1998} discussed several ways in which to bring model continuum
spectra in closer agreement with the observed spectrum in UX~UMa.  One
possibility, first discussed by \citet{long1994} with respect to discrepant
disk model fits to the spectrum of the NL, IX~Vel, is that the hottest,
inner annuli of the disk are not present, causing the disk spectrum to be
less blue than predicted by a standard disk model. The presence of 29~s
oscillations in the optical and UV light in UX~UMa is an inner disk
phenomenon and may be indicative of disrupted accretion at the innermost
annuli \citep{knigge1998b}, but the oscillations are not always present and
we do not detect them our observations. Another way to bring model spectra
into agreement with observed spectra is to posit a second continuum source:
\citet{knigge1998} showed that the addition of optically thin recombination
emission can flatten both the slope of the continuum from red to blue and
fill in the Balmer jump.  The optically thin continuum emission may
originate from the accretion disk chromosphere; further modeling will be
needed to determine if this picture is consistent with theoretical
understanding of the properties of the disk-wind transition region.

The WD may also be a significant source of UV continuum emission.  If so,
its presence will worsen the discrepancy between the models and the
observed continuum in UX~UMa \citep{knigge1998}.  \citet{baptista1995}
concluded from their analysis of UV light curves that the WD is seen in the
UV. They identified a signal in the derivative of the eclipse light curve
from which they determined contact points for the WD eclipse ingress and
egress.  They also determined the flux of the WD at 1600~\AA, 4.8~mJy
(\expu{5.625}{-14}{$\flam$}), which was 25\% of the continuum flux at the
time of their observation. From their identification of the WD in the UV
eclipse and a distance estimate of 345~pc obtained from their optical and
UV eclipse maps, \citet{baptista1995} determined the WD diameter and a
temperature: 52,000~K if the full WD is visible, or 70,000~K if the lower
hemisphere of the WD is obscured by the disk. They combined their measure
of the duration of WD eclipse with an empirical ZAMS mass-radius relation
and limits on $K_{1}$ to determine the binary parameters for UX~UMa,
including the mass ratio, inclination, binary separation, and masses and
radial velocities of the primary and the secondary. Their derived
parameters have become the standard adopted values for UX~UMa.

\begin{figure*}
\psfig{file=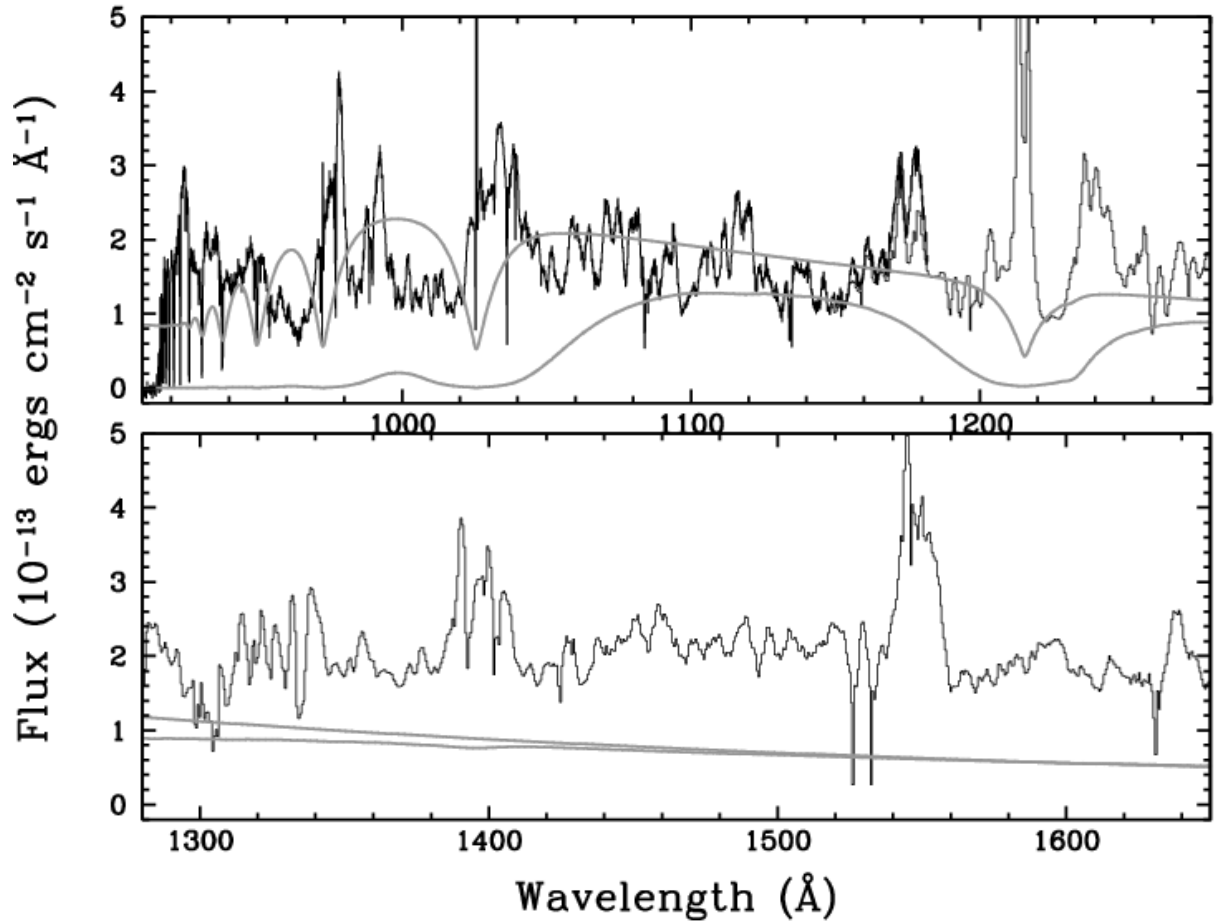,height=5in,angle=-90}
\figcaption[f9.eps]{The FUV and UV spectra of UX~UMa compared to
model white dwarf spectra.  The time-averaged FUV spectrum of UX~UMa
from the 2001 March FUSE observations is shown as a bold black
line. Plotted in a thin black line is the 1996 November GHRS UV
spectrum of UX~UMa.  Overplotted in gray are two model DA WD spectra
scaled to have fluxes of 4.8~mJy (\expu{5.625}{-14}{$\flam$}) at
1600~\AA.  Both WD models assume $\log g = 8$.  The upper model has a
temperature of 52,000~K, while the lower model has a temperature of
20,000~K. \label{fig_wd}}
\end{figure*}

A simple extension of the spectrum of a WD with the flux and temperature
found by \citet{baptista1995} to FUV wavelengths indicates a problem with
their analysis, however. This is illustrated in Figure~\ref{fig_wd}, in
which we compare model WD spectra to the 2001 FUV and (for reference) the
1996 November UV spectra of UX~UMa. The DA WD models shown were created
using TLUSTY and SYNSPEC \citep{hubeny1988,hubeny1994,hubeny1995}, and have
$\log g = 8$.  A WD with a 1600~\AA\ flux of 4.8~mJy
(\expu{5.625}{-14}{$\flam$}) and a temperature of 52,000~K exceeds the
observed flux at nearly every wavelength in the FUV; only in regions of
strong FUV line emission does the model even fall below the observed
flux. At its peak near 1000~\AA, the model WD spectrum exceeds the observed
flux by 60\%.  A 70,000~K WD (the Baptista et al.\ temperature if half the
WD is obscured by the disk) will deviate even more severely from the
observed FUV spectrum.

We examined the possibility that the WD flux found by
\citet{baptista1995} is correct but that their temperature is too
high.  The second WD model in Figure~\ref{fig_wd} shows that a cooler
WD with a 4.8~mJy flux at 1600~\AA\ is equally unlikely, however.  The
model shown is for a 20,000~K WD, which is the hottest WD of that UV
flux consistent with the observed FUV spectrum. The problem with a
cool WD with substantial UV flux is evident: its spectral shape is too
variable with wavelength to be consistent with the observed
spectrum. Moreover, such a WD would need to have a radius of
\expu{2.3}{9}{cm} -- \expu{3.7}{9}{cm} (for the 216 -- 345~pc distance
range to UX~UMa) to emit the flux shown, substantially larger than the
\expu{9.8}{8}{cm} radius found by Baptista et al.

We conclude, therefore, that the WD in UX~UMa is unlikely to have the high
flux value determined by Baptista et al.  A strong change in the WD flux
over time is also unlikely.  The 1600~\AA\ flux in 1993 (the time of the
Baptista et al.\ observations) is roughly the same as the flux in the 1996
HST spectrum.  A comparison of the Baptista et al. WD model to that of the
1996 spectrum in Figure~\ref{fig_wd} shows that the model begins to
overshoot the observed flux around 1200~\AA, even before an accretion disk
contribution to the FUV flux is taken into account.  This inconsistency
between the derived WD parameters and observed FUV spectra calls into
question whether the features identified by Baptista et al.\ in the UV
eclipse were actually the ingress and egress of the WD.  The S/N of the UV
eclipse light curve used by Baptista et al.\ to identify the WD was not
high, and we consider their identification of the WD contact points in the
derivative light curve as insecure. As all subsequent binary parameters
found by \citet{baptista1995} depend on the accurate identification of the
WD in the eclipse, we also consider their determination of the WD
temperature, WD radius, primary and secondary masses, and inclination
unreliable. The derived parameters themselves indicate a problem: a mass
ratio of 1, as found by Baptista et al., is theoretically inconsistent with
stable mass transfer, for which there is ample and unambiguous evidence in
UX~UMa.

\begin{figure*}
\psfig{file=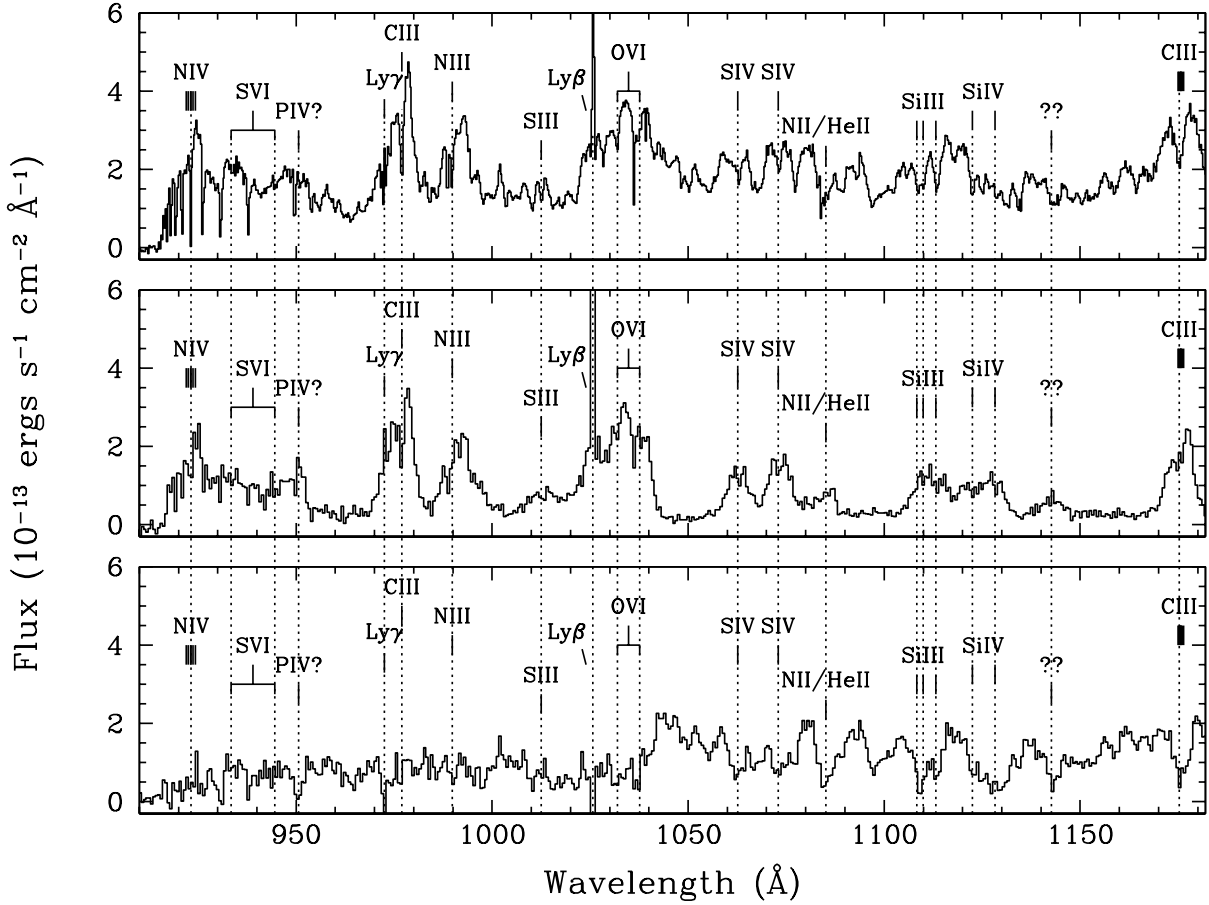,height=5in,angle=-90}
\figcaption[f2.eps]{The out of eclipse and mid-eclipse spectra of UX~UMa
and the spectrum of the eclipsed light.  The upper panel shows the out of
eclipse spectrum in black, binned to 0.1~\AA\ resolution. This spectrum is
the average spectrum pre- and post-eclipse (orbital phases 0.85 -- 0.95 and
0.05 -- 0.15). Prominent lines and lines present in eclipse are labelled and
marked with dotted lines.  The middle panel shows the spectrum of the
central 200~s (0.0125~cycles) of the eclipse binned to 0.6~\AA\
resolution. The lower panel shows the spectrum of the eclipsed light, which
is the difference spectrum between the out of eclipse and mid-eclipse
spectra (both were binned to 0.6~\AA\ before subtraction).
\label{fig_eclipse}}
\end{figure*}

\begin{figure*}
\psfig{file=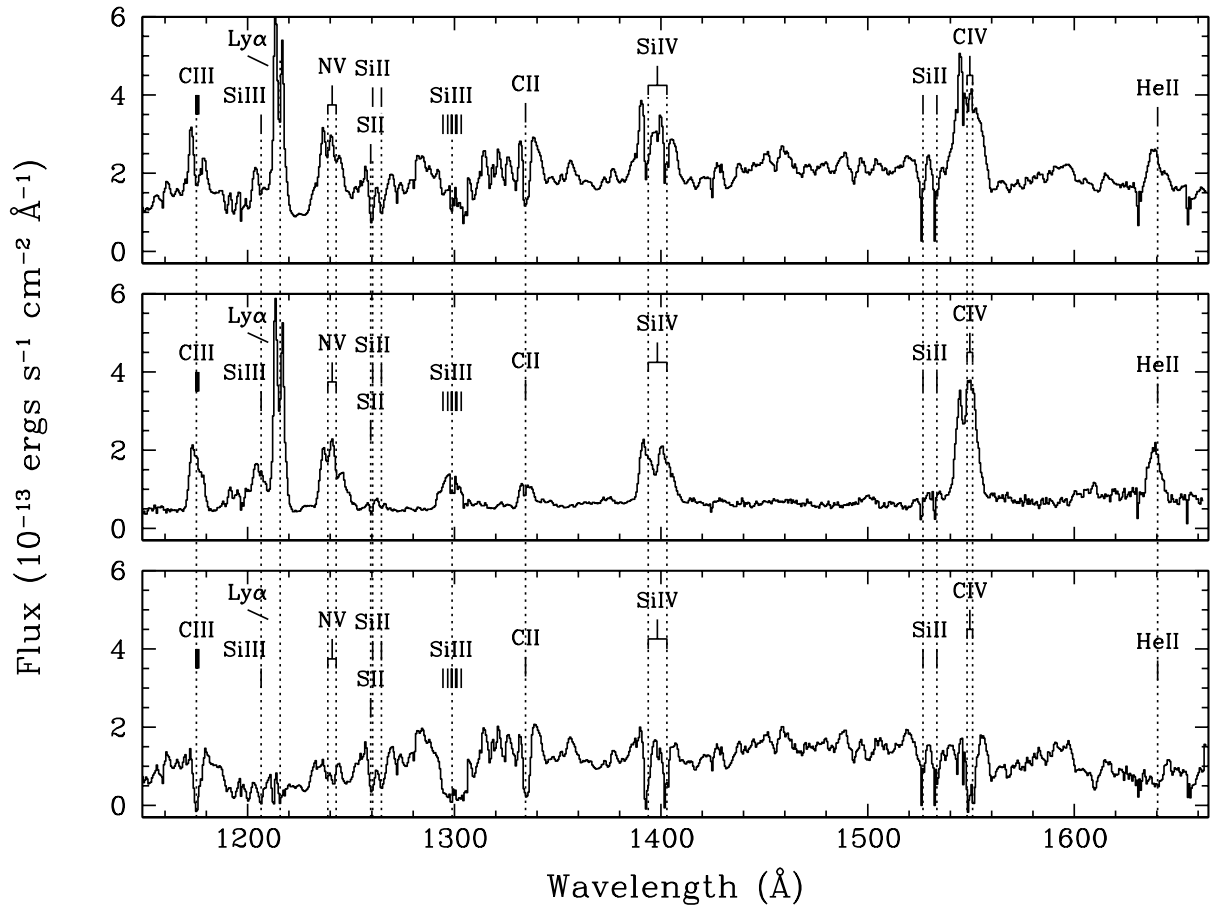,height=5in,angle=-90}
\figcaption[f7.eps]{The time-averaged and mid-eclipse spectra
from 1996 GHRS observations of UX~UMa (upper and middle panels) and
the spectrum of the eclipsed light (lower panel). The labelling of the
panels is the same as in Figure~\ref{fig_eclipse}. All three spectra
are shown at their maximum resolution, 0.6~\AA. \label{fig_eclipse_hst}}
\end{figure*}

Although the WD does not have the large flux contribution found by
Baptista et al., it may still be an important source of flux in
the FUV.  We do not see obvious WD eclipse features in the FUV
high time resolution light curve, however.  Figure~\ref{fig_eclc}
shows a slight notch in eclipse egress near phase 0.02, but that
feature is only present in one of the two eclipses we observed
(the individual eclipses in broadband FUV light can be seen in
Figure~\ref{fig_contlc}). We can set limits on the WD contribution
based on the spectrum of the eclipsed light, shown in the bottom
panel of Figure~\ref{fig_eclipse}. Since the WD is fully eclipsed,
the flux in the eclipsed light spectrum gives an upper limit to
the WD contribution.  If we assume a standard WD mass $M_{WD}
\simeq 0.6 \msun$, a corresponding radius of order $R_{WD} \simeq
8.7\times10^{8}$~cm (using the mass-radius relation from Nauenberg
1972), and a maximum distance of 345~pc, we find that the maximum
temperature of the WD cannot exceed 40,000~K if the WD flux is to
fall below the observed FUV eclipsed flux.  At this temperature
and projected area, the WD would be responsible for all of the
eclipsed flux, which is unlikely, considering that most of the
inner accretion disk and the chromosphere are also eclipsed, and
no clear-cut WD eclipse features can be seen in the FUV light
curve.  If we assume therefore that the WD is only responsible for
50\% of the FUV eclipsed flux, its temperature would drop to
25,000~K, and a 20\% WD contribution would predict a cool WD of
20,000~K.

Theoretically, half of the energy of mass accretion is emitted in the
boundary layer at or near the WD surface; for the typical accretion
rates of UX~UMa, this indicates that the surface of the WD (or some
part thereof) should be quite hot ($\sim$100,000~K).  If the WD is
hot, it must have a small emitting area to be consistent with the
eclipsed FUV flux. If fully visible, a 50,000~K WD must have a radius
equal to or smaller than \expu{3.1}{8}{cm}, assuming the minimum
distance to UX~UMa of 216~pc.  A 70,000~K WD would have to have a
radius $R_{WD} \leq 2.4\times10^8$~cm for its flux to be smaller than
the observed FUV eclipsed flux.  Both of these values correspond to a
massive WD (M$_{WD} > 1.2 \msun$) if a standard WD mass-radius
relation holds for for UX~UMa. Observationally, a hot BL/WD has not
been observed in UX~UMa, whose X-ray luminosity
($10^{30}$~ergs~s$^{-1}$ for a 340~pc distance) is too low to be
consistent with the theoretically predicted BL emission
\citep{wood1995}.  This may indicate that we are not seeing the WD in
UX~UMa at all due to shrouding or occultation by the outer
disk. Occultation of the WD has been demonstrated in the the NL DW~UMa
($\imath = 82\arcdeg$) when a low state in the disk exposed the
normally hidden WD \citep{knigge2000}.  A clear detection of the
eclipse of the WD would rule out the occultation hypothesis.  Ideally,
a search for the contact points of the WD would be conducted using
multiple (at least 10) UV observations (with FUSE or HST) of the
eclipse to decrease the effects of flickering and variability in the
shape of the eclipse.

\section{Conclusions} \label{sec_conc}

In this manuscript, we presented the observational characteristics of
the first high resolution time series FUV spectroscopy of the
eclipsing NL, UX~UMa.  Our main results are as follows:

\noindent1.\ The time-averaged FUV spectrum of UX~UMa is dominated by
line emission from numerous, overlapping transitions of \ion{H}{1},
\ion{C}{3}, \ion{N}{3}, \ion{N}{4}, \ion{S}{3}, \ion{S}{4},
\ion{S}{6}, \ion{Si}{3}, \ion{Si}{4} and \ion{O}{6}, and additional
unidentified, blended features.  There is no region in the
time-averaged spectrum in which the continuum alone can be discerned.

\noindent2.\ The emission lines have double-peaked profiles, with
a narrow (FWHM $\simeq$ 600~km~s$^{-1}$) central absorption
reversal at the rest wavelength of each line.  In the strongest
lines --- \ion{C}{3} $\lambda$977~\AA, \ion{N}{3}
$\lambda$992~\AA, the \ion{N}{4} blend at 920--924~\AA, and the
doublet lines of \ion{O}{6} $\lambda\lambda$1032,1038~\AA\ --- the
red emission peak is enhanced relative to the blue peak.

\noindent3.\ The mid-eclipse spectrum is much less complex than the
time-averaged spectrum.  Most of the weaker spectral features are
occulted in eclipse, exposing a flat continuum with the prominent
emission lines superposed.  The central absorption components of these
lines are eclipsed.  The spectrum of the eclipsed light is a mix of
continuum and line emission, with the line features primarily in
absorption. Several lines appear in emission only at mid-eclipse, when
the strong absorption components of the lines are occulted. Archival
HST spectra of UX~UMa show similar morphology and behavior during
eclipse.

\noindent4.\ Over the full FUV range (910--1182~\AA), the mid-eclipse
flux is 40\% below the uneclipsed level.  For regions of the spectrum
away from the prominent emission lines, the mid-eclipse flux is on
average 82\% below the uneclipsed level.  The eclipse depth does not
change significantly at longer UV wavelengths: the mid-eclipse flux is
77\% of the mean out of eclipse flux in the continuum-dominated
regions of the HST (1149 -- 1660~\AA) spectra.  The full width at half
depth of the eclipse away from the prominent emission lines is roughly
the same over the FUV and UV as well.

\noindent5.\ The high time resolution (1~s) light curve of the FUV
observation shows strong flickering on time scales of several
minutes. We do not detect a periodicity associated with the flickers
and we do not detect the 29~s oscillations seen in some previous
optical and UV observations.  Spectra of the flickering peaks and
minima show that the flickering occurs in regions near the plane of
the disk (i.e., the regions that are eclipsed); the continuum fluctuates
but the prominent emission lines do not flicker.

\noindent6.\ There is an orbital variation in the FUV light curve: the
phase-binned flux around orbital phase 0.55 dips to a minimum 25\%
below the flux at other phases outside of eclipse.  The light curve
dip occurs at roughly the same orbital phase as but is shallower than
the dip seen in HST observations of UX~UMa.  Spectra in the light
curve dip, averaged over two orbits, show lower EWs for the line
absorption features than spectra taken outside the dip.  The same
behavior was seen in the 1996 HST data.

\noindent7.\ We have detected a wavelength shift in the FUV spectrum phased
on the orbital period.  The amplitude of the shift is 70~km~s$^{-1}$ over
broad regions of the FUV but may range as high as 140 -- 200~km~s$^{-1}$
for the absorption components of the lines.  The wavelength shifts lag the
phasing of the photometric ephemeris by $\sim21\arcdeg$.

\noindent8.\ The WD parameters found by \citet{baptista1995} from HST
UV observations requires a FUV flux from the WD in excess of the
observed FUV flux.  A WD with a radius \expu{8}{8}{cm} --
\expu{9.5}{8}{cm} (corresponding to a mass range 0.7 -- 0.47~$\msun$)
must be cooler than 40,000~K to remain below the observed FUV flux.
Alternatively, the WD can be hotter but smaller in radius.

\noindent9.\ A curve of growth analysis of interstellar absorption
lines in the FUV spectrum of UX~UMa (see Appendix~\ref{sec_cog}) gives
a range of $\log N{_H} = 20.3^{+0.5}_{-0.8}$ for the column density
along the line of sight to UX~UMa.  

\acknowledgements{Based on observations made with the NASA-CNES-CSA Far
Ultraviolet Spectroscopic Explorer. FUSE is operated for NASA by the Johns
Hopkins University under NASA contract NAS5-32985.  Our thanks to the FUSE
operations team for their work in scheduling and acquiring the observations
as well as their prompt responses concerning the use of the CALFUSE
pipeline to conduct time-series spectroscopy.  We also gratefully
acknowledge the financial support from NASA through grant NAG5-10381.}

\pagebreak

\begin{figure*}
\plotone{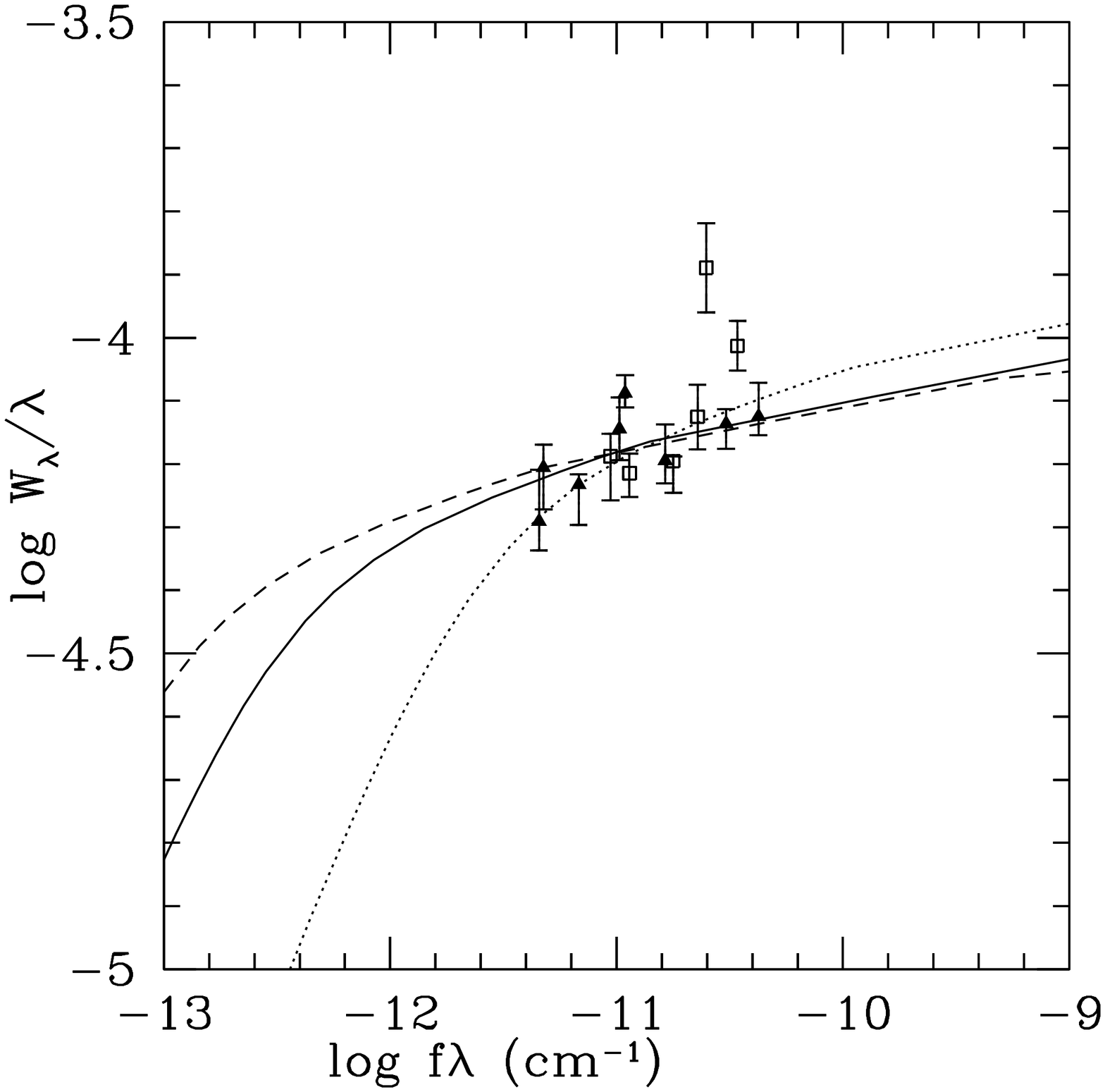}
\figcaption[f10.eps]{Curve of growth for the sight line to UX~UMa.  O~I
lines are shown as filled triangles and N~I lines as open squares.
Both species have been adjusted for their gas-phase interstellar
abundances with respect to hydrogen.  The absissca is plotted for
$\log f \lambda$ rather than the usual $\log N f \lambda$ so that
growth curves of multiple column densities can be shown on the same
plot.  The parameters of the best-fit growth curve, shown as a solid
line, are $N_{H} = 2.13\times10^{20}$ and $b = 4.5$~km~s$^{-1}$.  The
dashed lines shows a curve for $N_{H} = 5.72\times10^{20}$~cm$^{-2}$
and $b = 4.1$~km~s$^{-1}$, while the dotted line shows the curve for
$N_{H} = 3.5\times10^{19}$~cm$^{-2}$ and $b =
5.9$~km~s$^{-1}$\label{fig_cog}}
\end{figure*}

\appendix
\section{A Curve of Growth Analysis of the Interstellar Lines in the
Spectrum of UX~UMa} \label{sec_cog}

The time-averaged spectrum of UX~UMa shows narrow interstellar
(IS) absorption lines.  Identified lines and their EWs are given
in Table~\ref{tab_ism}. The EWs of the lines are measured averages
from time-averaged spectra of each of the four optical channels,
giving two or four measurements of each line, depending on the
wavelength region. The IS lines are from \ion{H}{1} and metals;
there is no molecular hydrogen absorption in the spectrum. The
lack of H$_{2}$ absorption is somewhat unusual since most FUSE
spectra, especially for targets outside the local ISM, show
interstellar H$_{2}$, and distance estimates place UX~UMa at least
216~pc away, and probably closer to 345~pc away (see Baptista et
al. 1995 and sources therein). Non-local sight lines without
molecular hydrogen absorption do exist, however (see Shull et al.\
2000, for example), and UX~UMa lies on a sight line directed out
of the galactic plane ($l = 107\fdg0$, $b = 63\fdg8$).

At least two distinct velocity components contribute to interstellar
absorption along the line of sight to UX~UMa.  The IS \ion{H}{1} lines
have FWHM $\sim$0.36~\AA, about twice that seen in the spectra of
nearby CVs observed by FUSE (Froning et al.\ 2001; Long et al. 2002,
in preparation). The strong, ionized metal lines --- \ion{C}{3}
$\lambda$977~\AA\, the blend of \ion{N}{3} $\lambda$989.8~\AA\ and
\ion{Si}{3} $\lambda$989.9~\AA, \ion{C}{2} $\lambda$1036~\AA, and
\ion{N}{2} $\lambda1084$~\AA\ --- have two components, one near zero
velocity and a second at a velocity of -60~km~s$^{-1}$ (the resolution
element in the individual channel spectra is $\sim$25~km~s$^{-1}$). We
examined archival STIS high-resolution (R = 0.075~\AA) spectra in the
vicinity of \ion{C}{4} $\lambda\lambda$1548,1552~\AA\ (the spectra are
unpublished).  The interstellar component of the nearby \ion{Si}{2}
$\lambda$1527.71~\AA\ line also shows two absorption components
separated by 60~km~s$^{-1}$.  The IS EWs given in Table~\ref{tab_ism}
for the FUSE spectra are integrated over the full IS absorption
profile of each line.

A measure of the absorbing column is of general interest for analysis
of UX~UMa, particularly for X-ray and EUV observations, so we
undertook a determination of the hydrogen column density along the
line of sight to UX~UMa using a curve of growth analysis.  The higher
order Lyman series lines of \ion{H}{1} cannot be used to directly
determine $N_{H}$ because they are all saturated.  Instead, we used
measurements of unblended \ion{O}{1} and \ion{N}{1} IS lines and
assumed standard IS abundances relative to hydrogen for these species
(see Mauche, Raymond \& C\'{o}rdova 1988 and Froning et al.\ 2001 for
other applications of this method). The gas-phase interstellar
abundances of oxygen and nitrogen have been well determined
\citep{meyer1997,meyer1998} and are reliable tracers of hydrogen
within several hundred parsecs of the sun (but see Moos et al.\ 2002
concerning nitrogen, which may be less reliable than oxygen as a
tracer of hydrogen in the LISM).

Unfortunately, none of the measured metal lines are completely
optically thin, so a direct measurement of the column density
independent of the broadening parameter governing the line widths is
not possible.  The metals also cluster in a small range of optical
depths and do not tightly constrain the column density.  The best fit
to the \ion{O}{1} and \ion{N}{1} line EWs gives $N_{H} =
2.13\times10^{20}$~cm$^{-2}$ and $b$ = 4.5~km~s$^{-1}$ for the
broadening parameter ($\chi^{2} = 32$ with 12 degrees of freedom).
The best fit for \ion{O}{1} alone gives the same result.  Fits within
the 95\% confidence interval for $N_{H}$ (assuming Gaussian
statistics) can be obtained for $N_{H}$ down to
$3.50\times10^{19}$~cm$^{-2}$ ($b = 5.9$~km~s$^{-1}$). The fit
deviates more rapidly from the data points for larger column densities
as the growth curve becomes too flat to match the slope in the
observed points, leading to an upper estimate of the column density
(for a 95\% confidence interval) of $N_{H} =
5.72\times10^{20}$~cm$^{-2}$ ($b = 4.1$~km~s$^{-1}$).  The resulting
range for the column density for the line of sight to UX~UMa is $\log
N{_H} = 20.3^{+0.5}_{-0.8}$. The best fit and the upper and lower
statistical limits are shown in Figure~\ref{fig_cog}.

The curve of growth analysis assumes a single cloud with a single
broadening parameter along the line of sight.  The broad widths and
large EWs of the IS \ion{H}{1} lines show that multiple clouds must
contribute to absorption in neutral hydrogen. If placed on the curve
of growth, the \ion{H}{1} lines are described by a broadening
parameter of $b \simeq$ 24~km~s$^{-1}$. This value is implausibly
large --- if dominated by thermal broadening, $b$ = 24~km~s$^{-1}$
implies a thermal temperature of 34,600~K! The growth curve
over-estimates the broadening parameter in \ion{H}{1} with a single
cloud; two or more clouds with offset velocities will have smaller,
more physical, broadening parameters in each cloud and be consistent
with observed EWs. The more pertinent question concerns the
reliability of the column density derived above, based on \ion{O}{1}
and \ion{N}{1} line measurements. In the absence of high resolution
spectroscopy or complicated modeling of the sight line to UX~UMa
beyond the scope of this manuscript, we cannot know the number and
parameters of IS absorption components along the sight line, and any
number of scenarios could be devised to make the true \ion{H}{1}
column density larger or smaller than the range indicated by the curve
of growth.  There is reason to believe that the column density range
given above is a reasonable estimation for the line of sight to
UX~UMa, however. \citet{jenkins1986} demonstrated that the curve of
growth method can give column densities close to the true value even
when multiple components with well-separated or partially overlapping
velocities contribute to the absorption.  This is true even when
individual components are very optically thick, as long as the
distribution of cloud parameters is smooth and no single or few highly
saturated components dominate the absorption. There is no evidence
from the IS spectrum of UX~UMa that any divergent component dominates
the absorption, so we conclude that, given the current data, the range
of $\log N{_H} = 20.3^{+0.5}_{-0.8}$ is the best estimate for the
column density along the line of sight to UX~UMa.

\clearpage
\begin{deluxetable}{ccccccc}
\tablecaption{Observation Summary\label{tab_obs}} \tablewidth{0pt}
\tablecolumns{7} \tablehead{ \colhead{Observation} & \colhead{Date
(UT)} & \colhead{Start (UT)} & \colhead{$\Phi_{Start}$} &
\colhead{End (UT)} & \colhead{$\Phi_{End}$} & \colhead{t$_{obs}$
(s)} } \startdata
1\dotfill & 2001 March 23 & 22:42:36 & 0.27 & 13:39:55 & 3.62\tablenotemark{a} & 20,038 \\
2\dotfill & 2001 March 24 & 15:20:02 & 0.80 & 02:59:41 & 3.35\tablenotemark{a} & 19,814 \\
\enddata
\tablenotetext{a}{Each observation was acquired over about three
binary orbits.}
\end{deluxetable}

\clearpage
\begin{deluxetable}{lccc}
\tablecaption{Intersteller Lines in the FUV Spectrum of UX~UMa\label{tab_ism}}
\tablewidth{0pt}
\tablecolumns{4}
\tablehead{
\colhead{Species} & \colhead{$\lambda_{lab}$\tablenotemark{a}}
& \colhead{$f$\tablenotemark{a}} & \colhead{EW} \\
\colhead{} & \colhead{(\protect\AA)} & \colhead{} & \colhead{(\protect\AA)}}
\startdata
H~\sc{i} &  915.329 & 0.000385 & 0.3908 \\
H~\sc{i} &  915.824 & 0.000468 & 0.3901 \\
H~\sc{i} &  916.429 & 0.000577 & 0.4184 \\
H~\sc{i} &  917.181 & 0.0007226 & 0.4226 \\
H~\sc{i} &  918.129 & 0.0009213 & 0.4346 \\
H~\sc{i} &  919.351 & 0.00120 & 0.4215  \\
H~\sc{i} &  920.963 & 0.001605 & 0.4290 \\
H~\sc{i} &  923.150 & 0.002216 & 0.4125 \\
O~\sc{i} &  924.950 & 0.001540 & 0.0473 \\
H~\sc{i} &  926.226 & 0.003183 & 0.426 \\
O~\sc{i} &  929.517 & 0.002295 & 0.0544 \\
H~\sc{i} &  930.748 & 0.004816 & 0.4600 \\
O~\sc{i} &  936.630 & 0.003650 & 0.0765 \\
H~\sc{i} &  937.803 & 0.007804 & 0.4451 \\
O~\sc{i} &  948.686 & 0.005420 & 0.0606 \\
H~\sc{i} &  949.743 & 0.01394  & 0.4349 \\
O~\sc{i} &  950.885 & 0.001570 & 0.0592 \\
N~\sc{i} &  953.415 & 0.01314  & 0.0619 \\
N~\sc{i} &  953.655 & 0.02492  & 0.0608 \\
N~\sc{i} &  953.970 & 0.03479  & 0.1231 \\
N~\sc{i} &  963.990 & 0.0148   & 0.0534 \\
N~\sc{i} &  964.626 & 0.0094   & 0.0607 \\
N~\sc{i} &  965.041 & 0.00675  & 0.0676 \\
O~\sc{i} &  971.738 & 0.01367  & 0.0729 \\
H~\sc{i} &  972.537 & 0.0290   & 0.3131\tablenotemark{b} \\
O~\sc{i} &  976.448 & 0.003310 & 0.0700 \\
C~\sc{iii} & 977.020 & 0.7620  & 0.2482 \\
O~\sc{i} &  988.578 & 0.000553 & 0.2021 \\
O~\sc{i} &  988.655 & 0.0083   & \nodata\tablenotemark{c} \\
O~\sc{i} &  988.773 & 0.04650  & \nodata\tablenotemark{c} \\
N~\sc{iii} & 989.799 & 0.1066  & 0.1892 \\
Si~\sc{ii} & 989.873 & 0.1330  & \nodata\tablenotemark{c} \\
H~\sc{i} &  1025.722 & 0.0791  & \nodata\tablenotemark{b} \\
O~\sc{i} &  1025.762 & 0.01705 & \nodata\tablenotemark{b,c} \\
C~\sc{ii} & 1036.337 & 0.1231  & 0.2602 \\
O~\sc{i} &  1039.230 & 0.009200 & 0.0760 \\
Ar~\sc{i} & 1048.220 & 0.26275  & 0.0404 \\
Ar~\sc{i} & 1066.660 & 0.06652 &  0.0236 \\
N~\sc{ii} & 1083.990 &  0.1031 &  0.2357 \\
N~\sc{i} & 1134.165 &  0.01342 & 0.0692 \\
N~\sc{i} & 1134.415 &  0.02683 & 0.085 \\
N~\sc{i} & 1134.980 &  0.04023 & 0.1102 \\
\enddata
\tablenotetext{a}{Wavelengths and oscillator strengths are taken from
Morton (1991). The oscillator strengths of some lines have been
updated by D.~C. Morton since the time of the previous publication;
these were obtained from Jenkins et al.\ (2000).}
\tablenotetext{b}{Line is contaminated by terrestrial airglow.}
\tablenotetext{c}{Line is blended with the previous transition.}
\end{deluxetable}

\clearpage
\begin{deluxetable}{cccccc}
\tablecaption{Lines in the FUV Mid-Eclipse Spectrum of UX~UMa\label{tab_eclipse}}
\tablewidth{0pt}
\tablecolumns{6}
\tablehead{
\colhead{Species} & \colhead{$\lambda_{lab}$} &
\colhead{$\lambda_{obs}$} & \colhead{EW} & \colhead{FWHM}
& \colhead{FWZI} \\
\colhead{} & \colhead{(\protect\AA)} & \colhead{(\protect\AA)}
& \colhead{(\protect\AA)} & \colhead{(km s$^{-1}$)}
& \colhead{(km s$^{-1}$)} }
\startdata
N~\sc{iv} & 923.2\tablenotemark{a} & 924.5 & \nodata\tablenotemark{b} & \nodata\tablenotemark{b} & \nodata\tablenotemark{b} \\
P~\sc{iv} & 950.7 & 950.0 & -50 & 2100 & 2900 \\
C~\sc{iii} & 977.0 & 976.7 & -90 & 2600 & 4900 \\
N~\sc{iii} & 989.8\tablenotemark{a} & 992.1 & -69 & 2550 & 5300 \\
S~\sc{iii} & 1012.5 & 1012.8 & -30 & 3400 & 5300\tablenotemark{b} \\
H~\sc{i} & 1025.7 & 1025.0 & -25 & 1600 & 4100\tablenotemark{b} \\
O~\sc{vi} & 1031.9,1037.62 & 1034.4 & -160 & 3300 & 5800 \\
S~\sc{iv} & 1062.7 & 1062.9 & -58 & 2050 & 4900 \\
S~\sc{iv} & 1073.0 & 1073.2 & -58 & 2200 & 4750 \\
N~\sc{ii}/He~\sc{ii} & 1085.0 & 1085.1 & -19 & 1750 & 3700\tablenotemark{b} \\
Si~\sc{iii} & 1108.4\tablenotemark{a} & 1110.9 & -113\tablenotemark{c} & 2700 & 5500\tablenotemark{b} \\
Si~\sc{iv} & 1122.5,1128.3 & 1124.8 & \nodata & 3600 & 6000\tablenotemark{b} \\
?? & \nodata & 1142.7 & -61.2 & 2300 & 5200 \\
C~\sc{iii} & 1175.3\tablenotemark{a} & 1176.0 & -88.0 & 1950 &
4500\tablenotemark{b} \\
\enddata
\tablenotetext{a}{This feature is a blend of multiple transitions of
the same ion.  The laboratory wavelength given is for the transition
with the largest oscillator strength in the blend.}
\tablenotetext{b}{Measurement uncertain due to line blending or the
detector edge.}
\tablenotetext{c}{Equivalent width for this line and the following together.}
\end{deluxetable}

\end{document}